\documentclass[final]{IEEEtran}
\usepackage{amsthm,amssymb,graphicx,multirow,amsmath,color,amsfonts}
\usepackage[update,prepend]{epstopdf}
\usepackage[noadjust]{cite}
\usepackage[latin1, utf8]{inputenc}
\usepackage{tikz}
\usepackage{bbm} 
\usepackage{pdfpages}
\usepackage{tabulary}
\usepackage{multirow}
\usepackage{comment}
\usepackage{subfigure}
\usepackage{float}
\usepackage[titlenumbered,ruled]{algorithm2e}
\usepackage[noend]{algpseudocode}
\DeclareUnicodeCharacter{0177}{\^y}

\def\nb0{{\mathbf{0}}}
\def\nb1{{\mathbf{1}}}

\newtheorem{lemma}{Lemma}

\newtheorem{definition}{Definition}

\newtheorem{theorem}{Theorem}

\allowdisplaybreaks 
\begin{document}
\title{Unveiling Passive and Active EMF Exposure in Large-Scale Cellular Networks}
\author{
	Yujie Qin, Mustafa A. Kishk, {\em Member, IEEE}, Ahmed Elzanaty, {\em Senior Member, IEEE}, Luca Chiaraviglio, {\em Senior Member, IEEE},  and Mohamed-Slim Alouini, {\em Fellow, IEEE}
	\thanks{Yujie Qin and Mohamed-Slim Alouini are with Computer, Electrical and Mathematical Sciences and Engineering (CEMSE) Division, King Abdullah University of Science and Technology (KAUST), Thuwal, 23955-6900, Saudi Arabia (e-mail: yujie.qin@kaust.edu.sa; slim.alouini@kaust.edu.sa).}
	\thanks{Mustafa Kishk is with the Department of Electronic Engineering, Maynooth University, Maynooth, W23 F2H6, Ireland (e-mail: mustafa.kishk@mu.ie).}
	\thanks{A. Elzanaty is with the 5GIC \& 6GIC, Institute for Communication Systems (ICS), University of Surrey, Guildford, GU2 7XH, United Kingdom (e-mail: a.elzanaty@surrey.ac.uk).}
	\thanks{L. Chiaraviglio is with the Department of Electronic Engineering,	Universita degli Studi di Roma Tor Vergata, 00133 Rome, Italy and Consorzio Nazionale Interuniversitario per le Telecomunicazioni (CNIT), Parma, Italy (e-mail: luca.chiaraviglio@uniroma2.it).}
}

\maketitle
\begin{abstract}
	With the development of fifth-generation (5G) networks, the number of user equipments (UE) increases dramatically.  However, the potential health risks from electromagnetic fields (EMF) tend to be a public concern. Generally, EMF exposure-related analysis mainly considers the passive exposure from base stations (BSs) and active exposure that results from the user's personal devices while communicating. However, the passive radiation that is generated by nearby devices of other users is typically ignored. In fact, with the increase in the density of UE, their passive exposure to human bodies can no longer be ignored. In this work, we propose a stochastic geometry framework to analyze the EMF exposure from active and passive radiation sources. In particular, considering a typical user, we account for their exposure to EMF from BSs, their own UE, and other UE. We derive the distribution of the Exposure index (EI) and the coverage probability for two typical models for spatial distributions of UE, i.e.,  \textit{i)} a Poisson point process (PPP); \textit{ii)} a Matern cluster process. Also, we show the trade-off between the EMF exposure and the coverage probability. Our numerical results suggest that the passive exposure from other users is non-negligible compared to the exposure from BSs when user density is $10^2$ times higher than BS density, and non-negligible compared to active exposure from the user's own UE when user density is $10^5$  times the BS density. 
\end{abstract}
\begin{IEEEkeywords}
	Electric and magnetic fields exposure; uplink transmission; Poisson point process; Matern cluster process; passive and active exposure.
\end{IEEEkeywords}
\section{Introduction}
The fifth-generation of cellular networks (5G) has been designed to guarantee low delays and high throughput, accommodate high user density, meet the goal of wide coverage,  and connect the unconnected regions \cite{andrews2014will, ITU2010}. 
In fact, the number of mobile users has increased from $3.6$ billion to $5.2$ billion from 2014 to 2020 \cite{Tillekeratne:20}. 
Despite much research effort and advanced communication techniques that have been accomplished to design rate and energy-efficient networks, developing networks with reduced electromagnetic field (EMF) exposure is still an open problem that concerns the public \cite{chiaraviglio2021dense}. 


Generally, EMF exposure is classified as passive and active exposure, in which passive exposure is commonly composed of exposure from base stations (BSs) and active exposure resulting from the user's personal devices.
Based on the guidelines provided by International Telecommunication Union (ITU) \cite{itu2004guidance}, International Commission on Non-Ionizing Radiation Protection (ICNIRP) \cite{international1998guidelines}, and Federal Communications Commission (FCC) \cite{us2020human}, each country has its regulations regarding the safety limits for EMF exposure and some restrictions on BSs deployment \cite{7739164}. These guidelines and regulations are mainly developed to guarantee that the temperature elevation of the  exposed tissues is within a safe limit \cite{foster2018thermal,belpomme2018thermal}. 

Surprisingly, most of the concerns about EMF exposure are from the deployment of BSs, even if the emissions from mobile phones present a significant component as the UE antennas are much closer to the human body \cite{sambo2014survey}. In this regard, it is a fundamental task  to guarantee that the EMF exposure from all the components (e.g., BS and UE) in future networks is below the acceptable safe levels \cite{leitgeb2016potential,chiaraviglio2021health}. 


Motivated by the aforementioned importance of designing EMF-efficient networks and considering the exponential growth in the number of UE, in this work, we propose a stochastic geometry framework to analyze the EMF exposure of the whole network. In particular, considering a typical user, we account for their exposure to EMF from BSs, their own UE, and other UE. Specifically, we use tools from stochastic geometry to model the locations of UE and BSs and study the passive and active exposure.

\subsection{Related Work}
Literature related to this work can be categorized into: (i) experimental EMF exposure assessment, (ii) analytical EMF exposure assessment and design. A brief discussion on related works in each of these categories is discussed in the following lines.

{\em EMF exposure experimental assessment.} Authors in \cite{kim2020human} investigated the EMF exposure in 5G systems compared to the EMF exposure measured in previous generations. 
Authors in \cite{9521570} measured the EMF from a BS by using an integrated approach and performing a large set of measurements over the territory.   A spectrum analyzer-based measurement methodology was proposed in \cite{aerts2019situ} to evaluate the time-averaged instantaneous exposure. Some other approaches to assess 5G EMF exposure were provided in \cite{colombi2020analysis,carciofi2020rf}.  For instance, authors in \cite{colombi2020analysis} analyzed the EMF for commercial 5G networks but considered transmit power monitoring of a set of 5G BSs. Authors in \cite{carciofi2020rf} pointed out the need for dynamic monitoring of EMF exposure and proposed a dynamic measurement approach for carrying out rapid extensive monitoring of radio-frequency EMF which can be used to characterize the dynamics of EMF in 5G systems. 

{\em Analytical EMF exposure assessment and design.}
For EMF-aware cellular system design, a resource allocation scheme was developed in \cite{sambo2016electromagnetic} to minimize the EMF exposure. Rate-splitting multiple access (RSMA)-based with EMF constraints was investigated in \cite{jiang2023rate} where the uplink energy efficiency is enhanced while limiting the EMF exposure under a certain level. A reconfigurable intelligent surface (RIS)-assisted network was analyzed in \cite{ibraiwish2021emf}  to  minimize the EMF exposure of users while maintaining a minimum target quality of service. In addition, authors in \cite{subhash2022optimal} proposed to maximize  the minimum SINR in RIS networks while limiting the EMF exposure of users considering  statistical  channel state information (CSI). 

For analytical EMF assessment, authors in \cite{thors2017time} presented a statistical model, which is based on modeling the expectation of the statistically conservative fraction of the total power contributing to the EMF exposure within an arbitrary beam, to estimate the time-averaged realistic maximum power levels for the assessment of EMF exposure using massive MIMO. In \cite{baracca2018statistical}, authors proposed another statistic model, which is based on a three-dimensional spatial channel model, for modeling the 5G EMF exposure of massive MIMO systems. As for the effect of massive MIMO on the EMF exposure, authors in \cite{chiaraviglio2021pencil} revealed that the pencil beamforming may be beneficial for the network throughput and reducing the EMF level. A novel  approach was proposed in \cite{chiaraviglio2023dominance} for EMF analysis  by comparing the radio BS EMF and EMF generated by 5G smartphones. 

While previous related works only analyze the EMF exposure from a relatively small number of BSs, large-scale network EMF analysis is essential for 5G network design. In \cite{chiaraviglio2018planning}, the problem of 5G BS locations planning under EMF exposure limits is considered. A stochastic geometry approach to EMF exposure modeling was proposed in \cite{9462948} and the author validated the model using experimental data. Manhattan Poisson line process (MPLP) was used in \cite{wiame2023joint} to derive the joint distributions of data rate and EMF exposure.
A stochastic geometry-based analysis was used in \cite{gontier2023joint} and authors analyzed the impact of the network parameters and discussed the trade-off between coverage and EMF exposure.
Besides, a stochastic geometry-based analysis of EMF in sub-$6$ GHz and mmWave networks was proposed in \cite{muhammad2021stochastic}, in which the locations of mmWave BSs and sub-$6$ GHz BSs are modeled by Poisson point processes (PPPs).
In \cite{10047969}, Poisson hole process is used to capture the impact of having restricted areas where BSs are not allowed to be deployed. Next, the downlink, uplink, and joint downlink \& uplink exposure induced by the radiation from BSs and personal UE are analyzed.

Different from the existing literature, which mainly neglects the EMF exposure from active uplink users,
our work analyzes the total exposure from all network components including the passive EMF exposure from active uplink users. More details on the contributions of this paper are provided next.

\subsection{Contributions}
This paper investigates the EMF exposure of a network composed of BSs and UEs. Different from existing literature, which only focuses on the impact of the EMF passive exposure from BSs and active exposure from the user's personal equipment, we capture the passive exposure which is composed of all active uplink users and BSs. We study the impact of system parameters such as user density, uplink path-loss compensation factor, and the number of antennas on the EMF exposure.
The main contributions of this work are summarized as follows.

\begin{itemize}
	\item We propose a stochastic geometry-based framework to study the EMF exposure of two types of users:  (i) passive users and (ii) active users, in which the passive user does not participate in the communication process (not using UE), and the active user participates in uplink communication. In particular, we consider the exposure from all sources, both BSs and active uplink UE. 
	
	\item In order to capture the scenario of spatially-clustered UE, which is commonly found in different use cases such as in airports or schools we model the locations of UE using MCP. To the best of the authors' knowledge, this is the first work to address the influence of UE spatial-clustering on EMF exposure. The inhomogeneitiy in the spatial distribution of UE in the case of MCP introduces some mathematical challenges, specially in computing total passive uplink EMF exposure, which we carefully handle in this paper. We also provide the analysis in the case of PPP to establish meaningful comparisons. 
	\item Using tools from stochastic geometry, we are able to compute the mean and cumulative density function (CDF) of exposure index (EI, which is a performance metric to characterize the level of EMF exposure formally defined in Definition~\ref{Exposure Index}) under PPP and MCP models, respectively. 
\end{itemize}

\section{System Model}
We are interested in studying the impact of user distributions, user densities, BS density, and uplink transmission on EMF exposure.  In this regard, we consider a wireless cellular network where the locations of BSs are modeled by a PPP, denoted by $\Phi_{\rm b}$ with density $\lambda_{\rm b}$, we consider two types of user distributions, PPP and MCP, respectively \cite{9153823,afshang2017nearest,9986039}.
\begin{figure*}
	\centering
	\subfigure[]{\includegraphics[width = 0.9\columnwidth]{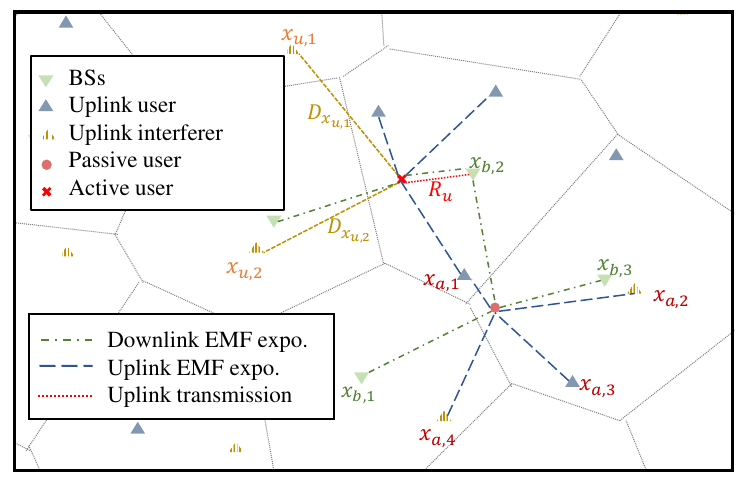}}	
	\subfigure[]{\includegraphics[width = 0.9\columnwidth]{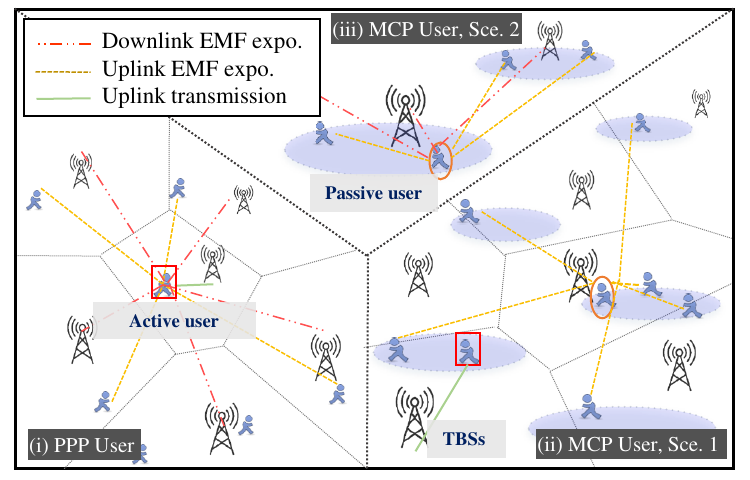}}
	\caption{Illustration of system model: \textbf{(a)} Passive and active user, and \textbf{(b-i)} PPP user, \textbf{(b-ii)} MCP user, Scenario 1, and \textbf{(b-iii)} MCP user, Scenario 2.}
	\label{fig_sys}
\end{figure*}
\begin{table*}\caption{Table of Notations}\label{table_par_ana}
	\centering
	\begin{center}
		\resizebox{1.6\columnwidth}{!}{
			\renewcommand{\arraystretch}{1}
			\begin{tabular}{ {c} | {c} }
				\hline
				\hline
				\textbf{Notation}&\textbf{Description}  \\ \hline
				$\Phi_{\rm b}$, $\Phi_{\rm u}$ & Point set of the locations of BSs,  locations of PPP users  \\ \hline
				$x_{\rm b}$, $x_{\rm u}$, $x_a$ & Locations of BSs,  locations of interfering uplink users, locations of active uplink users  \\ \hline
				$R_{\rm u}$, $D_{\{\cdot\}}$ & Distance to the serving BS, distance between two locations\\\hline
				$\Phi_{\rm c}$, $\Phi_{\rm cu}$ & Locations of user cluster centers,  locations of  MCP users   \\ \hline
				$\lambda_{\rm b}$, $\lambda_{\rm u}$, $\lambda_{\rm c}$, $\lambda_{\rm cu}$ & Density of BSs, density of PPP users, density of  user clusters, density of users in a cluster \\ \hline
				$\Phi_{\rm ue}$, $\Phi_{\rm ue}^{\prime}$, $\Phi_{\rm ue}^{''}$ & Locations of all uplink users, locations of active uplink users, locations of interfering uplink users   \\ \hline
				$\eta$, $\alpha$, $\beta$ & Power inversion control factor, path-loss of communication, path-loss of exposure\\ \hline
				$\rho_{\rm b}$, $\rho_{\rm u}$, $p_{\rm max}$ & Transmit power of BSs, minimum transmit power of UEs, maximum transmit power of UEs\\ \hline
				$W_{\rm b}$, $W_{\rm b,a}$ & EMF exposure from BSs (passive user, PPP user model), EMF exposure from BSs (active user, PPP user model)\\ \hline
				$W_{\rm b}^{\prime}$, $W_{\rm b,a}^{\prime}$ & EMF exposure from BSs (passive user, MCP user model), EMF exposure from BSs (active user, MCP user model)\\ \hline
				$W_{\rm u}$, $W_{\rm u,\{1,2\}}$ & EMF exposure from uplink users (PPP user model), EMF exposure from uplink users (MCP user model, passive user, Sce. 1 \& 2)\\ \hline
				$W_{\rm u,a,\{1,2\}}$ & EMF exposure from uplink users (MCP user model, active user, Sce. 1 \& 2)\\ \hline
				${\rm SAR}_{\rm ul}$, 	${\rm SAR}_{\rm dl}$ & Reference induced SAR in uplink,  Reference induced SAR in downlink\\ \hline
				${\rm EI}_{\rm p}$, 	${\rm EI}_{p,\{1,2\}}$, ${\rm EI}_{\rm a}$ & Exposure Index (passive user, PPP user model), Exposure Index (MCP users, Sec. 1 \& 2), Exposure Index (active user)\\ \hline
				${\rm EI}_{\rm bs}$, 	${\rm EI}_{\rm ul,u}$, ${\rm EI}_{\rm ul,tr}$ & Exposure Index of BSs, Exposure Index of active uplink users, Exposure Index of the uplink transmission of active user\\ \hline
				$P_{\rm cov}$, 	$I$, $\sigma^2$ & Coverage probability, interference, noise power
				\\\hline\hline
		\end{tabular}}
	\end{center}
\end{table*}
In this work, we compute the EMF exposure considering two types of users: (i) a passive user and (ii) an active user (active uplink user), defined below.

\begin{definition}[Passive User]
	Passive user denotes a user who is not an uplink user. Therefore, the EMF exposure of the passive user is composed of the exposure from (i) all BSs, due to the downlink transmission between BSs and users, and (ii) all active uplink users. Besides, we assume that the passive user is located at the origin (stationarity of PPP \cite{haenggi2012stochastic}).
\end{definition}
\begin{definition}[Active User]
\label{def_activeuser}
	Active user denotes an active uplink user. Therefore, the EMF exposure of the active user is composed of the exposure from (i) all BSs, (ii) all other active uplink users, and (iii) its uplink transmission. Without loss of generality, we randomly select one active uplink user as our reference user.
\end{definition}
The system model is shown in Fig.~\ref{fig_sys}  and related notations are explained in Table \ref{table_par_ana} and the definitions below.


\begin{definition}[PPP User  model]
	At places where users are scattered uniformly on the whole plane, we model the locations of users by a PPP denoted by $\Phi_{\rm u}$ with density $\lambda_{\rm u}$.
\end{definition}
\begin{definition}[MCP User model]
	At places where users are spatially clustered, we model the locations of users by a MCP. Particularly, we consider two scenarios: Scenario 1: the locations of user cluster centers are independent to BSs and modeled by a PPP, denoted by $\Phi_{\rm c}$, with density $\lambda_{\rm c}$, and  Scenario 2: user clusters are centered at BSs, which means that users are scattered around the BSs. For both scenarios, users are uniformly distributed in the user clusters with radius $r_{\rm c}$, and let $\Phi_{\rm cu}$ denote the locations of all cluster users,  and $\lambda_{\rm cu}$ denote the user density of a user cluster.
\end{definition}
To simplify the notations, we use $\Phi_{\rm ue}$ to present the locations of all uplink users, e.g., in the case of PPP user  model $\Phi_{\rm ue} = \Phi_{\rm u}$ and in the case of MCP  user  model $\Phi_{\rm ue} = \Phi_{\rm cu}$. Let $\Phi_{\rm ue}^{'}$ denote the locations of the active uplink users which is obtained from an independent thinning of $\Phi_{\rm u}$ with an active probability $p_a$ \cite{haenggi2012stochastic}. 

In the uplink transmission, users associate with their nearest BSs and the association regions of the users with the BSs form a Poisson-Voronoi (PV) tessellation 
\cite{9444343,6786498}. In addition, we assume that the BSs allocate the resources to avoid inter-cell interference. Hence, from the perspective of a certain resource block and a certain PV cell, only one uplink user is active and the interfering users can only be from other cells. Let $\Phi_{\rm ue}^{''}$ be the locations of the interfering users for the uplink transmission; hence, $\Phi_{\rm ue}^{''}$ is a subset of $\Phi_{\rm ue}^{'}$. However, all active uplink users contribute to EMF exposure regardless of their different resource blocks and association cells. 

\subsection{Communication Model}

We first introduce the communication model of the considered system. We consider BSs equipt with omnidirectional antenna with antenna gain $G_b$, and for the uplink transmission, the received power at the BS is given by
\begin{align}
	p_r = G_b\, p_{\rm u}\, H\,  R_{\rm u}^{-\alpha},
\end{align}
in which the variable $H$ models the small-scale Rayleigh fading of the channel and follows the exponential distribution with a mean of unity. We assume that the inversion power control technology considered in  \cite{10024838} is adopted. From \cite{10024838}, for a given $R_{\rm u}=r_{\rm u}$, the truncated transmission power of a user  is given by
\begin{align}
	p_\textrm{u} =\left\{
	\begin{aligned}
		\rho_{\rm u}\, r_{\rm u}^{\alpha\,\eta}&,\quad 1\leq r_{\rm u} <r_0,\\
		p_{\rm max}&, \quad r_{\rm u} \geq r_0,\\
	\end{aligned}
	\right.
\end{align}
in which $r_{\rm u}$ denotes the distance between the user and the serving BS, $\rho_{\rm u}$ is the minimum transmit power for the adopted uplink power control technique, $\alpha$ is the path-loss for calculating SINR and $\eta$ is the path-loss compensation factor, $p_{\rm max}$ denotes the maximum transmit power of a mobile user, and $r_0 = (p_{\rm max}/\rho_{\rm u})^{1/\alpha\eta}$.

Recall that we randomly select one active uplink user as our reference user and its serving BS as tagged BS. Generally, coverage probability is defined as the probability that the reference user is successfully served, which denotes the event that the SINR of the related link is above a predefined threshold. 
\begin{definition}[Uplink Coverage Probability]\label{def_cov}
	The uplink coverage probability is defined as
	\begin{align}
		P_{\rm cov}(\gamma) = \mathbb{P}({\rm SINR}>\gamma),
	\end{align}
	in which ${\rm SINR}$ denotes the signal-to-interference-plus-noise ratio at the serving BS, $I$ is the aggregate interference, and $\gamma$ is the required SINR threshold,
	\begin{align}
		{\rm SINR} &= \frac{ p_r}{\sigma^{ 2}+I},\quad
		I = \sum_{x\in\Phi_{\rm ue}^{''}} p_{u,x} H_{x}  D_{x}^{-\alpha},
	\end{align}
	in which $\sigma^{ 2}$ is the normalized uplink transmission noise power: $\sigma^{ 2} = \frac{\sigma^{\prime 2}}{l_{\rm u}}$, $l_{\rm u}$ the path-loss at the reference distance $d_0 = 1$ m, $l_{\rm u}= (\frac{c/f_{\rm u}}{4\pi d_0})^2$ and $\sigma^2$ is the noise power, $p_{u,x}$ denotes the transmit power, $H_{x}$ denotes the channel fading of user $x\in\Phi_{\rm ue}^{''}$, and $D_{x}$ is the distance between the user $x_{\rm u}$ and the tagged BS.
\end{definition}

\subsection{EMF Exposure}
EI is the performance metric defined to characterize individual exposure and it is built from the aggregation of exposure from different sources and situations. In the proposed system, the exposure sources are composed of BSs and active uplink users and the personal mobile of the active user. To compute the EI of the passive user and the active user, we first analyze the EMF exposure from the aforementioned sources.

Let $p_{\rm b}$ and $p_o$ be the received power from a BS and UE, respectively,
\begin{align}
	p_{\rm b}(D_{x}) &= \rho_{\rm b}\, G_b\, H_{x}\, D_{x}^{-\beta},\quad
	p_o( D_{x} ) = p_{u,{x}}\, H_{x}\,  D_{x}^{-\beta},\nonumber
\end{align}
where $\rho_{\rm b}$ is the transmit power of the BSs, $ D_{\{\cdot\}}$, $H_{\{\cdot\}}$ present the distance and channel fading, respectively, and $\beta$ denotes the EMF path-loss, where $\beta>2$ is assumed to guarantee the convergence of the integration. Consequently, the incident power densities from BSs and other UEs\footnote{Note that when we analyze the $W_{\rm u}$ for the active user defined in Def. \ref{def_activeuser}, $\Phi_{\rm ue}^{'}$ denotes the locations of other active UEs except the one we analyze.}, $W_{\rm b}$ and $W_{\rm u}$, are respectively given by
\begin{align}
	W_{\rm b} &=\sum_{x\in\Phi_{\rm b}}\frac{ \rho_{\rm b}\, G_b\,H_{x}}{4\pi}D_{x}^{-\beta},\nonumber\\
	W_{\rm u} &= \sum_{x\in\Phi_{\rm ue}^{'}}\frac{ p_{u,{x}}\, H_{x}\, }{4\pi }D_{x}^{-\beta},
\end{align}
where $p_{u,{x}}$ is the truncated transmit power for the user located at $x$, $x\in\Phi_{\rm b}$. Recall that the EMF exposure of the passive user is composed of the exposure from all BSs and all active uplink users and the EMF exposure of the active user is composed of the exposure from all BSs, other active uplink users, and its uplink transmission. Therefore, the EI of the passive user and active user is formally defined as follows.

\begin{definition}[Exposure Index]\label{def_EI}
	\label{Exposure Index}
	The EI ${\rm EI}_{\rm p}$ of the passive user is defined as
	\begin{align}
		{\rm EI}_{\rm p} = {\rm EI}_{\rm bs}+{\rm EI}_{\rm ul,u},\label{eq_EI_p}
	\end{align}
	in which ${\rm EI}_{\rm bs}$ and  ${\rm EI}_{\rm ul,u}$ are the EI of BSs and active uplink users, respectively, and given by
	\begin{align}
		{\rm EI}_{\rm bs} &= {\rm SAR}_{\rm dl}\,W_{\rm b},\quad{\rm EI}_{\rm ul,u} = {\rm SAR}_{\rm dl}\,W_{\rm u}, \nonumber
	\end{align}
	where  ${\rm SAR}_{\rm dl}$  is the reference induced SAR in downlink when the received power density from BSs is unitary.
	
	Similarly, the EI of the active user is defined as
	\begin{align}
		{\rm EI}_{\rm a} = {\rm EI}_{\rm bs}+{\rm EI}_{\rm ul,u}+ {\rm EI}_{\rm ul,tr},\label{eq_EI_a}
	\end{align}
	in which $ {\rm EI}_{\rm ul,tr}$ is the EI of uplink transmission of the active user
	\begin{align}
		{\rm EI}_{\rm ul,tr} &= {\rm SAR}_{\rm ul}\,p_{\rm u},\nonumber
	\end{align}
	where ${\rm SAR}_{\rm ul}$ is the reference induced SAR in uplink when the transmit power from user equipment is unity, respectively.
\end{definition}

\section{EI Analysis of PPP  User  Model}
In this section, we provide the analysis of EI in the case that users are PPP distributed. To do so, we first compute the PDF of the transmit power and then obtain the Laplace transform of $W_{\rm b}$ and $W_{\rm u}$, respectively. Since the truncated path-loss inversion power control is used, the transmit power is a mixed random variable and its PDF is given in the following lemma.
\begin{lemma}[Distribution of the Transmit Power]
The PDF of $p_{\rm u}$ is given by
\begin{align}
	&f_{p_{\rm u}}(p) =
	\frac{2\pi\lambda_{\rm b}}{\alpha\eta\rho_{\rm u}^{\frac{2}{\alpha\eta}}}	p^{\frac{2}{\alpha\eta}-1}\exp\bigg(-\pi\lambda_{\rm b}\bigg(\frac{p}{\rho_{\rm u}}\bigg)^{\frac{2}{\alpha\eta}}\bigg)\nonumber\\
 &+
	\exp\bigg(-\pi\lambda_{\rm b}\bigg(\frac{p_{\rm max}}{\rho_{\rm u}}\bigg)^{\frac{2}{\alpha\eta}}\bigg)\delta_{p_{\rm max}}(p),\quad \rho_{\rm u}<p\leq p_{\rm max},
\end{align}
in which $\delta_{p_{\rm max}}(p)$ is an impulse at $p_{\rm max}$ and satisfies $\int_{0}^{\infty}\delta_{p_{\rm max}}(p){\rm d}p=1$.
\end{lemma}
\begin{IEEEproof}
This Lemma follows by using the PDF of the contact distance of a PPP.
\end{IEEEproof}
It is difficult to compute the CDF of ${\rm EI}_{\rm p}$ and ${\rm EI}_{\rm a}$ directly. Alternatively, we can use the Gil-Pelaez theorem which requires the characteristic function or Laplace transform of ${\rm EI}_{\rm p}$ and ${\rm EI}_{\rm a}$. Hence, in the following lemma, we first derive the Laplace transform of the $W_{\rm b}$ and $W_{\rm u}$, then the Laplace transform of ${\rm EI}_{\rm p}$ and ${\rm EI}_{\rm a}$ can be obtained by the multiplication of the Laplace transform of $W_{\rm b}$ and $W_{\rm u}$.

\subsubsection{Passive User}
We first compute the Laplace transform of $W_{\rm b}$ and $W_{\rm u}$ for the passive user. Recall that the passive user is located at the origin, and the passive EMF exposure includes the exposure from all BSs and active uplink users.
\begin{lemma}[Laplace Transform of $W_{\rm u}$ and $W_{\rm b}$ of the Passive User]\label{lemma_WuWb_passive}
For the passive user, the Laplace transform of $W_{\rm u}$ and $W_{\rm b}$ is given by
	\begin{align}
	\mathcal{L}_{W_{\rm u}}(s) =& \exp\bigg(-2\pi\lambda_{\rm u} p_a\int_{0}^{\infty}\int_{\rho_u}^{p_{\rm max}}\nonumber\\
 &\bigg(\frac{f_{p_{\rm u}}(x)}{1+4\pi  (s x)^{-1} z^{\beta} }\bigg){\rm d}x z{\rm d}z\bigg),\nonumber\\
	\mathcal{L}_{W_{\rm b}}(s) = & \exp\bigg(-\lambda_{\rm b}\int_{0}^{\infty}\bigg(\frac{1}{1+4\pi  (s \rho_{\rm b} G_{\rm b})^{-1} z^{\beta}}\bigg) z{\rm d}z\bigg).\nonumber
\end{align}
\end{lemma}
\begin{IEEEproof}
The Laplace transform of $W_{\rm u}$ is given by
\begin{align}
	\mathcal{L}_{W_{\rm u}}(s) 
	&= \mathbb{E}_{\Phi_{\rm u}^{'},H,p_{\rm u}}\bigg[\prod_{x\in\Phi_{\rm u}^{'}}\exp\bigg(-s\frac{p_{\rm u} H ||x||^{-\beta}}{4\pi  }\bigg)\bigg]\nonumber\\
	&\stackrel{(a)}{=} \mathbb{E}_{\Phi_{\rm u}^{'},p_{\rm u}}\bigg[\prod_{x\in\Phi_{\rm u}^{'}}\bigg(\frac{1}{1+s p_{\rm u} ||x||^{-\beta}(4\pi  )^{-1}}\bigg)\bigg]\nonumber\\
	&\stackrel{(b)}{=} \exp\bigg(-2\pi\lambda_{\rm u} p_a\int_{0}^{\infty}\int_{\rho_u}^{p_{\rm max}}\bigg[1-\nonumber\\
 &\quad\bigg(\frac{1}{1+s p z^{-\beta}(4\pi  )^{-1}}\bigg)\bigg]f_{p_{\rm u}}(p){\rm d}p z{\rm d}z\bigg),
\end{align}
in which step (a) follows from using the MGF of exponential random variable and step (b) follows from using PGFL of PPP. The Laplace transform of $W_{\rm b}$ follows similar steps, thus omitted here.
\end{IEEEproof}

\begin{lemma}[Laplace Transform of ${\rm EI}_{\rm p}$]
The Laplace transform of ${\rm EI}_{\rm p}$ is given by
\begin{align}
	\mathcal{L}_{\rm EI_p}(s) &=  \mathcal{L}_{W_{\rm b}}(s\,{\rm SAR}_{\rm dl})\mathcal{L}_{W_{\rm u}}(s\,{\rm SAR}_{\rm dl}),
\end{align}
in which $\mathcal{L}_{W_{\rm b}}(s), \mathcal{L}_{W_{\rm u}}(s)$ are obtained in Lemma \ref{lemma_WuWb_passive}.
\end{lemma}
\begin{IEEEproof}
The Laplace transform of ${\rm EI}_{\rm p}$ is computed by
\begin{small}
	\begin{align}
		&\mathcal{L}_{\rm EI_p}(s) = \mathbb{E}_{\rm EI_p}[\exp(-s\, {\rm EI_p})]\nonumber\\
		&= \mathbb{E}_{W_{\rm u},W_{\rm b}}\bigg[\exp\bigg(-s\,{\rm SAR}_{\rm dl}(W_{\rm b}+W_{\rm u})\bigg)\bigg]\nonumber\\
		&\stackrel{(a)}{\approx}\mathbb{E}_{W_{\rm u}}\bigg[\exp\bigg(-s\,{\rm SAR}_{\rm dl}\,W_{\rm u}\bigg)\bigg]\mathbb{E}_{W_{\rm b}}\bigg[\exp\bigg(-s\,{\rm SAR}_{\rm dl}\,W_{\rm b}\bigg)\bigg]\label{EI_t_ppp_laplace}
	\end{align}
 \end{small}
in which step (a) follows from assuming $W_{\rm b}$ and $W_{\rm u}$ to be independent (note that this is an approximation\footnote{Note that the random variables $W_{\rm b}$ and $W_{\rm u}$ are generally not independent given that the adopted uplink power control technique makes the transmit powers of the users function of the locations of the BSs.} that shows good matching as evident in the numerical results section). The proof completes by substituting the Laplace transform of $W_{\rm u}$ and $W_{\rm b}$.
\end{IEEEproof}

Consequently, by using the Gil-Pelaez theorem \cite{gil1951note}, the CDF and mean  of ${\rm EI}_{\rm p}$ are given in the following theorem.
\begin{theorem}[CDF and Mean of ${\rm EI}_{\rm p}$] \label{theore_CDF_EI_t}
The CDF and mean of the total passive exposure, including exposure from BSs and active uplink users are, respectively, given by
\begin{align}
	F_{{\rm EI_p}}(w) &= \frac{1}{2}-\frac{1}{2j\pi}\int_{0}^{\infty}\frac{1}{t}[\exp(-jtw)\mathcal{L}_{\rm EI_p}(-jt)\nonumber\\
 &\quad-\exp(jtw)\mathcal{L}_{\rm EI_p}(jt)]{\rm d}t,\\
	\bar{{\rm EI}}_{\rm p} &= {\rm SAR_{\rm dl}}\frac{\lambda_{\rm u}\, p_a}{2 (\beta-2)} \bar{p}_{u,p}+{\rm SAR_{\rm dl}}\, \frac{\rho_{\rm b}G_{\rm b}\lambda_{\rm b}}{2 (\beta-2)},
\end{align}
in which $j$ is the imaginary unit, $\bar{p}_{u,p} = \int_{0}^{\infty}\max(p_{\rm max},\rho_{\rm u} r^{\eta\alpha})f_{R_u}(r){\rm d}r$ is the mean of uplink transmit power, where $f_{R_u}(r)=2\lambda_b\pi r \exp\left(-\pi\lambda_b r^2\right)$ is the PDF of PPP contact distance.
\end{theorem}
\begin{IEEEproof}
From Gil-Pelaez theorem \cite{gil1951note}, the CDF has the following relation with the characteristic function,
\begin{align}
	&F_{\rm EI_p}(w) = \frac{1}{2}-\frac{1}{\pi}\int_{0}^{\infty}\frac{1}{t}\Im(\exp(-jtw)\phi_{\rm EI_p}(t)) {\rm d}t\nonumber\\
	&=\frac{1}{2}-\frac{1}{2\pi}\int_{0}^{\infty}\frac{1}{jt}(\exp(-jtw)\phi_{\rm EI_p}(t)-\exp(jtw)\phi_{\rm EI_p}(-t)) {\rm d}t,
\end{align}
in which $\Im(\cdot)$ denotes the imaginary part of a complex number, and proof completes by substituting $\mathcal{L}_{\rm EI_p}(-jt)=\phi_{\rm EI_p}(t)$.

The mean of  ${\rm EI}_{\rm p}$ is given by
\begin{align}
	\mathbb{E}[{\rm EI_p}] &= {\rm SAR_{\rm dl}}(\mathbb{E}[W_{\rm u}]+\mathbb{E}[W_{\rm b}]),
\end{align}
in which 
\begin{align}
	\mathbb{E}[W_{\rm u}]&= \frac{\partial M_{W_{\rm u}}(s)}{\partial s}\biggm\vert_{s = 0} =\frac{\partial \mathcal{L}_{W_{\rm u}}(-s)}{\partial s}\biggm\vert_{s = 0} \nonumber\\
	&= \frac{2\pi\lambda_{\rm u} p_a}{4\pi  }\int_{0}^{\infty}\int_{0}^{\infty} \min(p_{\rm max},\rho_{\rm u} r^{\alpha\eta}) z^{-\beta+1} f_{R_{\rm u}}(r) {\rm d}r{\rm d}z\nonumber\\
	&=  \frac{\lambda_{\rm u} p_a}{2 (\beta-2)}\int_{0}^{\infty} \min(p_{\rm max},\rho_{\rm u} r^{\alpha\eta})f_{R_{\rm u}}(r) {\rm d}r,\\
	\mathbb{E}[W_{\rm b}]&= \frac{\partial \mathcal{L}_{W_{\rm b}}(-s)}{\partial s}\biggm\vert_{s = 0}= 2\pi\lambda_{\rm b}G_{\rm b}\int_{0}^{\infty} \rho_{\rm b}  (4\pi)^{-1} z^{-\beta+1} {\rm d}z \nonumber\\
	&= \rho_{\rm b}G_{\rm b} (4\pi)^{-1}\frac{2\pi\lambda_{\rm b}}{ (\beta-2)} .
\end{align}
\end{IEEEproof}

\subsubsection{Active User}
In the case of computing the EI of the active user, while the Laplace transform of $W_{\rm u}$ of the active user the same as the passive uer, the EMF exposure from BSs is correlated with the exposure from the uplink transmission of the active user, e.g., the distance to the nearest BS is the same as the distance considered while computing the uplink transmit power. 
Hence, in order to compute the Laplace transform of the total exposure of the active user, we need to compute the Laplace transform for each of the downlink exposure and the uplink self-exposure conditioned on the distance to the serving BS before taking the expectation over this distance.

 Hence, the Laplace transform of downlink exposure for active users, $W_{\rm b,a}$ (except the exposure from the nearest BS, which is $W_{b,0}$ and will be introduced later in the text), is computed as a function of the communication distance between the active user and its serving BS.
\begin{lemma}[Laplace Transform of $W_{\rm b,a}$ of the Active User]\label{lemma_WuWb_ref}
For the active user, the  Laplace transform of  $W_{\rm b,a}$ is given by
\begin{align}
	\mathcal{L}_{W_{\rm b,a}}(s,R_{\rm u}) &=  \exp\bigg(-\lambda_{\rm b}\int_{R_{\rm u}}^{\infty}\bigg(\frac{z^{-\beta+1}}{1+4\pi (s \rho_{\rm b} G_{\rm b})^{-1} }\bigg) {\rm d}z\bigg).\nonumber
\end{align}

\end{lemma}
Consequently, the Laplace transform of ${\rm EI}_{\rm a}$,  composed of the exposure from (i) the nearest BS, (ii) remaining BSs, (iii) active uplink users, and (iv) uplink transmission of the active user, is given in the following lemma.
\begin{lemma}[Laplace Transform of ${\rm EI}_{\rm a}$]
The Laplace transform of ${\rm EI}_{\rm a}$ conditioned on $R_{\rm u}$ is given by
\begin{align}
	\mathcal{L}_{\rm EI_a}(s,R_{\rm u})& = \mathcal{L}_{W_{\rm b},a}(s\,{\rm SAR}_{\rm dl},R_{\rm u})\,\mathcal{L}_{W_{\rm u}}(s\,{\rm SAR}_{\rm dl})\nonumber\\
	&\times\exp\bigg(-s\,{\rm SAR}_{\rm ul}\min(p_{\rm max},\rho_{\rm u} R_{\rm u}^{\alpha\eta})\bigg)\nonumber\\
	&\times \bigg(\frac{1}{1+s\,{\rm SAR}_{\rm dl}\, \rho_{\rm b}\, G_{\rm b}\, R_{\rm u}^{-\beta}(4\pi)^{-1}}\bigg).
\end{align}
\end{lemma}

In the following theorem, we provide the CDF of ${\rm EI}_{\rm a}$ as well as the mean of ${\rm EI}_{\rm a}$.
\begin{theorem}[CDF and Mean of ${\rm EI}_{\rm a}$] \label{theorem_CDF_Mean_EI_r_PPP}
The CDF of the total exposure, including exposure from BSs, active uplink users and the uplink transmission of the active user, is given by
\begin{align}
&F_{{\rm EI_a}}(w) = \frac{1}{2}-\frac{1}{2j\pi}\int_{0}^{\infty}\int_{0}^{\infty}\frac{1}{t}[\exp(-jtw)\mathcal{L}_{\rm EI_a}(-jt,r)\nonumber\\
 &-\exp(jtw)\mathcal{L}_{\rm EI_a}(jt,r)]f_{R_{\rm u}}(r){\rm d}r{\rm d}t,\nonumber
\end{align}
in which $f_{R_{\rm u}}(r) = 2\pi r\lambda_{\rm b}\exp(-\pi r^2)$.

The mean of ${\rm EI}_{\rm a}$ is given by
\begin{align}
\bar{{\rm EI}}_{\rm a} &= \int_{0}^{\infty}\bigg({\rm SAR_{\rm dl}}\, (4\pi)^{-1}\bigg(G_{\rm b}\rho_{\rm b} \,r^{-\beta}+\frac{2\pi\lambda_{\rm b}G_{\rm b}\,\rho_{\rm b} }{\beta-2}r^{2-\beta}\bigg)\nonumber\\
& +{\rm SAR_{\rm ul}}\,\min(p_{\rm max},\rho_{\rm u} r^{\alpha\eta})\bigg)f_{R_{\rm u}}(r) {\rm d}r\nonumber\\
 & +{\rm SAR_{\rm dl}}\,\frac{\lambda_{\rm u} p_a}{2(\beta-2)}\bar{p}_{u,p},
\end{align}
recall that $f_{R_{\rm u}}(r) = 2\pi r\lambda_{\rm b}\exp(-\pi \lambda_{\rm b} r^2)$.
\end{theorem}
\begin{IEEEproof}
Similar to the passive user case, the CDF of ${\rm EI}_{\rm a}$ can be obtained from the Laplace transform using Gil-Pelaez theorem \cite{gil1951note}.

For the mean of ${\rm EI}_{\rm a}$, recall that  ${\rm EI}_{\rm a}$ is composed of the exposure from (i) nearest BS, (ii) remaining BSs, (iii) active uplink users, and (iv) uplink transmission of the active user, in which (i), (iii) and (iv) are functions of the communication distance $R_{\rm u}$. Therefore, the mean is given by
\begin{align}
\mathbb{E}[{\rm EI_a}] &=  \mathbb{E}[{\rm SAR_{\rm dl}}\,W_{\rm u}]+\mathbb{E}[{\rm SAR_{\rm dl}}(W_{\rm b,a}+W_{\rm b,0})+{\rm SAR_{\rm ul}}\,p_{\rm u}],\nonumber
\end{align}
in which
\begin{align}
&\mathbb{E}[{\rm SAR}_{\rm dl}W_{\rm u}]	= {\rm SAR_{\rm dl}}\,\mathbb{E}\bigg[\sum_{x\in\Phi_{\rm ue}^{\prime}}\frac{ p_{u,x}  H_{x}}{4\pi}D_{x}^{-\beta}\bigg] \nonumber\\
&\stackrel{(a)}{=} \frac{ {\rm SAR_{\rm dl} } }{4\pi}\mathbb{E}\bigg[\sum_{x\in\Phi_{\rm ue}^{\prime}}p_{u,x}D_{x}^{-\beta}\bigg]\nonumber\\
&= \frac{ {\rm SAR_{\rm dl} } }{4\pi}\mathbb{E}\bigg[\sum_{x\in\Phi_{\rm ue}^{\prime}}\min(p_{\rm max},\rho_{\rm u} R_{u}^{\alpha\eta})D_{x}^{-\beta}\bigg]\nonumber\\
&\stackrel{(b)}{=} \frac{\lambda_{\rm u}\, {\rm SAR_{\rm dl} } }{2}  \int_{0}^{\infty}\frac{1}{z^{\beta}}z{\rm d}z\int_{0}^{\infty} \min(p_{\rm max},\rho_{\rm u} r^{\alpha\eta})f_{R_{\rm u}}(r) {\rm d}r \nonumber\\
&={\rm SAR_{\rm dl}}\frac{\lambda_{\rm u} p_a}{2(\beta-2)}\int_{0}^{\infty} \min(p_{\rm max},\rho_{\rm u} r^{\alpha\eta})f_{R_{\rm u}}(r) {\rm d}r\nonumber\\
&\stackrel{(c)}{=}  {\rm SAR_{\rm dl}}\frac{ \lambda_{\rm u} p_a}{2(\beta-2)}\bar{p}_{u,p},
\end{align}
where step (a) follows from the fact that all channel fading coefficients are independent and exponentially distributed random variables with unity mean, (similar to \cite[Lemma 2]{10066317} and \cite{8239664}), step (b) follows from Campbell's theorem \cite{haenggi2012stochastic} with the conversion from Cartesian to polar coordinates and the independence of $p_{u,x}$ and $D_x$, and step (c) follows from the definition of $\bar{p}_{u,p}$ given in the proof of Theorem~\ref{theore_CDF_EI_t},
\begin{align}
&\mathbb{E}[{\rm SAR_{\rm dl}}(W_{b,a}+W_{b,0})+{\rm SAR_{\rm ul}}\,\frac{H p_{\rm u}}{4\pi}]\nonumber\\
& = \mathbb{E}_{R_{\rm u}}\bigg[{\rm SAR_{\rm dl}}\mathbb{E}\bigg[\sum_{x\in\Phi_{\rm b}\setminus x_{b_0}}\frac{ \rho_{\rm b} G_{\rm b}H_{x}}{4\pi}D_{x}^{-\beta}\bigg]\nonumber\\
 &+{\rm SAR}_{\rm dl}\,\mathbb{E}\bigg[\frac{ \rho_{\rm b} G_{\rm b} H_{x_{b_0}}}{4\pi}R_{\rm u}^{-\beta}\bigg]+\frac{{\rm SAR}_{\rm ul}\,H}{4\pi}\min(p_{\rm max},\rho_{\rm u} R_{\rm u}^{\alpha\eta})\bigg]\nonumber\\
&=\mathbb{E}_{R_{\rm u}}\bigg[{\rm SAR_{\rm dl}}\frac{ \lambda_{\rm b}G_{\rm b}\rho_{\rm b}}{2(\beta-2)}R_{\rm u}^{2-\beta}\nonumber\\
 &+{\rm SAR_{\rm dl}}\frac{ \rho_{\rm b}G_{\rm b} }{4\pi}R_{\rm u}^{-\beta}+\frac{{\rm SAR_{\rm ul}}}{4\pi}\min(p_{\rm max},\rho_{\rm u} R_{\rm u}^{\alpha\eta})\bigg],
%
\end{align}
in which $x_{b_0}$ denotes the location of the serving BS, $b_0$ is the serving BS, and the proof completes by integrating over $R_{\rm u}$.
\end{IEEEproof}

%
%

\section{EI Analysis of MCP  User  Model}
\begin{figure}
\centering
\includegraphics[width = 0.75\columnwidth]{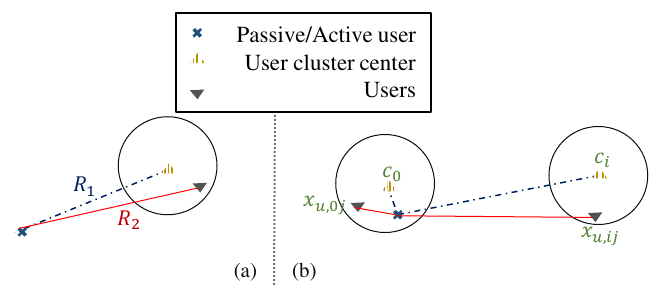}
\caption{Illustration of \textbf{(a)} conditional distance: the relation between $R_1$ and $R_2$, and \textbf{(b)} MCP system model: the locations of user cluster centers and users.}
\label{fig_sys_mcp}
\end{figure}

In this section, we provide the analysis of EI when the users are spatially clustered. To do so, we first compute the PDF of the distance between the reference user and an active uplink user from any cluster. Next, we obtain the Laplace transform of $W_{\rm b}$ and $W_{\rm u}$. Conditioning on the distance to the cluster center $R_1$, the distribution of the distance to the user in the cluster $R_2$, as shown in Fig.~\ref{fig_sys_mcp} (a), is given in the following lemma.

\begin{lemma}[Distance Distribution] Conditioned on the user cluster center being located at a distance $R_1$ from the user, the PDF of the distance to an active uplink user is given as in \cite{9986039} as follows.  For $R_1 < r_{\rm c}$, we have
\begin{align}
f_{R_2}(r_2|R_{1}) = \left\{
\begin{aligned}
	&\frac{2 r_2}{r_{\rm c}^{2}},\quad 0 \leq r_2 \leq r_{\rm c}-R_1,\\
	&\frac{2 r_2}{\pi r_{\rm c}^{2}}	\arccos\bigg(\frac{r_{2}^{2}+R_{1}^{2}-r_{\rm c}^{2}}{2 r_{2} R_{1}}\bigg)	,\nonumber\\
 & r_{\rm c}-R_1\leq r_2 \leq r_{\rm c}+R_{1},\\
\end{aligned}
\right.
\end{align}
while for $R_1 > r_{\rm c}$, we have
\begin{align}
f_{R_2}(r_2|R_{1}) &= 
\frac{2 r_2}{\pi r_{\rm c}^{2}}	\arccos\bigg(\frac{r_{2}^{2}+R_{1}^{2}-r_{\rm c}^{2}}{2 r_{2} R_{1}}\bigg)	,\nonumber\\
 & R_{1}-r_{\rm c}\leq r_2 \leq r_{\rm c}+R_{1},
\end{align}
and $f_{R_2}(r_2|R_{1}) = 0$, otherwise.
\end{lemma}

 In what follows, we compute the Laplace transform of $W_{\rm u}$. Note that in the case of MCP user model with PPP distributed BSs (Scenario 1), the Laplace transform of $W_{\rm b}$ and $W_{\rm b,a}$ are the same as the PPP user model since they are independent of the cluster centers. On the other hand, if BSs are located at the cluster centers (Scenario 2) $W_{\rm b}^{\prime} = W_{\rm b}+W_{\rm b,c}$ and $W_{\rm b,a}^{\prime} = W_{\rm b,a}+W_{\rm b,c}$, where $W_{\rm b,c}$ denotes the exposure of BS located at the reference user's cluster center. To compute the Laplace transform of $W_{\rm u}$, we first define the cluster point process.

A system illustration is shown in Fig.~\ref{fig_sys_mcp} (b). Let $\Phi_{\rm c}$ be the location of user cluster centers, modeled by a PPP with density $\lambda_{\rm c}$. With each point $c_i\in\Phi_{\rm c}$, associate with a point process $\Phi_{c_i,u}$. The point process $\Phi_{\rm c}$ is generally called the parent process and  $\Phi_{c_i,u}$ is called the offspring process. The union of offspring points processes is termed a cluster point process, denoted by $\Phi_{\rm cu} = \cup_{c_i\in\Phi_{\rm c}}c_i+\Phi_{c_i,u}$. In the case that the points in each point process, $x_{c_i,u_j}\in\Phi_{c_i,u}$, are uniformly distributed in the ball with radius $r_{\rm c}$ and center $c_i$, $B(c_i,r_{\rm c})$, $\Phi_{\rm cu}$ is so-called Matern cluster process (MCP). Consequently, the locations of each point in MCP are given by
\begin{align}
\Phi_{\rm cu} &= \biggl\{x_{u,ij}: c_i+x_{c_i,u_j}, c_i\in\Phi_{\rm c}, x_{c_i,u_j}\in\Phi_{c_i,u}, \{i,j\}\in\mathbb{N}\biggr\}.
\end{align}
Besides, conditioned on the passive user located at the origin and the passive user's cluster contains the origin, the Palm distribution of the mentioned MCP is given by $\Phi_{\rm cu}^{0} = \Phi_{\rm cu}\cup\{x_{u,0j}\}$ in which $\biggl\{x_{u,0j}: c_0+x_{c_0,u_j}, (0,0)\in B(c_0,r_{\rm c}), x_{c_0,u_j}\in\Phi_{c_0,u}, j\in\mathbb{N}\biggr\}$.
Similarly, let $p_a$ be the active probability of the uplink user and $\Phi_{\rm cu}^{0'}$ be the point set of the locations of the active uplink users. 
Recall that in the MCP user case, we consider two scenarios: Scenario 1: the locations of user clusters are independent of the locations of BSs, and Scenario 2:  user clusters are centered at BSs.  In Scenario 1, we assume that the active uplink users from the same cluster have the same transmit power (due to the relative small variance of transmit powers among users within the same cluster) and each user associates with its nearest BS. In Scenario 2, we assume that the user associates with the BS deployed at the cluster center. 

\subsubsection{Passive User}

The Laplace transform of $W_{\rm u,\{1,2\}}$ of the passive user, in which the subscript $\{1,2\}$ denotes Scenario 1 and Scenario 2, respectively, is given in the following lemma.

\begin{lemma}[Laplace Transform of $W_{\rm u}$, Scenario 1 \& 2] \label{lemma_Wu_passive_MCP}
In Scenario 1, the Laplace transform of $W_{\rm u}$ of the passive user is given by
\begin{align}
	&\mathcal{L}_{W_{\rm u,1}}(s)\approx  \exp\biggl\{-2\pi\lambda_{\rm c}\int_{0}^{\infty}1-\int_{0}^{\infty}\exp\bigg(-p_a\lambda_{\rm cu}\pi r_{\rm c}^{2}\nonumber\\
&\times\int_{0}^{r_{\rm c}+r_1}\frac{f_{R_2}(r| r_1)}{1+(s p(z))^{-1} (4\pi  ) r^{\beta}}{\rm d}r\bigg)f_{R_{\rm u}}(z){\rm d}zr_1{\rm d}r_1\biggr\} \nonumber\\
&\times \int_{0}^{r_{\rm c}}\exp\bigg(-p_a\lambda_{\rm cu}\pi r_{\rm c}^{2}\int_{0}^{r_{\rm c}+r_1}\frac{f_{R_2}(r| r_1)}{1+(s p(z))^{-1} (4\pi  ) r^{\beta}}{\rm d}r\bigg)\nonumber\\
&\times f_{R_{\rm u}}(z){\rm d}z \frac{2 r_1}{r_{\rm c}^{2}}{\rm d}r_1 ,
\end{align}
in which $
p(z) = 	\min(p_{\rm max},\rho_{\rm u} z^{\alpha\eta})$.

In Scenario 2, the Laplace transform of $W_{\rm u}$ of the passive user is given by
\begin{align}
	&\mathcal{L}_{W_{\rm u,2}}(s)\approx  \exp\biggl\{-2\pi\lambda_{\rm c}\int_{0}^{\infty}1- \exp\bigg(-p_a\lambda_{\rm cu}\pi r_{\rm c}^{2}\int_{0}^{r_{\rm c}+r_1}\nonumber\\
&\frac{f_{R_2}(r| r_1)}{1+(s \bar{p}_{u,m})^{-1} (4\pi  ) r^{\beta}}{\rm d}r\bigg) r_1{\rm d}r_1\biggr\}  \int_{0}^{r_{\rm c}}\int_{0}^{\infty}\exp\bigg(-p_a\lambda_{\rm cu}\pi r_{\rm c}^{2}\nonumber\\
&\int_{0}^{r_{\rm c}+r_1}\frac{f_{R_2}(r| r_1)}{1+(s \bar{p}_{u,m})^{-1} (4\pi  ) r^{\beta}}{\rm d}r\bigg)  \frac{2 r_1}{r_{\rm c}^{2}}{\rm d}r_1 ,
\end{align}
in which $
\bar{p}_{\rm u,m} = 	\int_{0}^{r_{\rm c}}\min(p_{\rm max},\rho_{\rm u} z^{\alpha\eta})\frac{2 z}{r_{\rm c}^{2}}{\rm d} z$.
\end{lemma}
\begin{IEEEproof}
See Appendix \ref{app_Wu_passive_MCP}.
\end{IEEEproof}

Similar to the PPP user model, we compute the CDF and mean of ${\rm EI}_{\rm p}$, which are given in the following theorem.
\begin{theorem}[CDF and Mean of ${\rm EI}_{\rm p}$, Scenario 1 \& 2] \label{theore_CDF_Mean_EI_t_MCP}
The CDF of the total passive exposure, including exposure from BSs and active uplink users is given by
\begin{align}
F_{{\rm EI_{\rm p,1}}}(w) &= \frac{1}{2}-\frac{1}{2j\pi}\int_{0}^{\infty}\frac{1}{t}[\exp(-jtw)\mathcal{L}_{\rm EI_{p,1}}(-jt)\nonumber\\
 &-\exp(jtw)\mathcal{L}_{\rm EI_{p,1}}(jt)]{\rm d}t,\nonumber\\
F_{{\rm EI_{\rm p,2}}}(w) &= \frac{1}{2}-\frac{1}{2j\pi}\int_{0}^{\infty}\int_{0}^{r_{\rm c}}\frac{1}{t}[\exp(-jtw)\mathcal{L}_{\rm EI_{p,2}}(-jt)\nonumber\\
 &-\exp(jtw)\mathcal{L}_{\rm EI_{p,2}}(jt)]f_{R_{u}^{\prime}}(r){\rm d}r{\rm d}t,\nonumber
\end{align}
in which
\begin{align}
\mathcal{L}_{\rm EI_{\rm p,\{1,2\}}}(s) &=  \mathcal{L}_{\{W_{\rm b},W_{\rm b}^{\prime}\}}(s\,{\rm SAR}_{\rm dl})\mathcal{L}_{W_{\rm u},\{1,2\}}(s\,{\rm SAR}_{\rm dl}),\nonumber\\
\mathcal{L}_{W_{\rm b}^{\prime}}(s) &=  \mathcal{L}_{W_{\rm b}}(s\,{\rm SAR}_{\rm dl})\exp\bigg(-s\,{\rm SAR}_{\rm dl}\frac{\rho_b G_{\rm b}}{4\pi  }R_{u}^{\prime-\beta}\bigg),
\end{align}
and $R_{u}^{\prime}$ has the PDF $f_{R_{u}^{\prime}}(r) = \frac{2r}{r_{\rm c}^2}$.

The mean of the total passive exposure, including exposure from BSs and active uplink users is given by
\begin{small}
\begin{align}
&\bar{{\rm EI}}_{\rm p,1} = {\rm SAR_{\rm dl}}\,\frac{\bar{p}_{u,p}\, \lambda_{\rm c}  p_a\lambda_{\rm cu}\pi r_{\rm c}^{2}}{2 } \int_{0}^{\infty}\int_{0}^{r_{\rm c}+r_1} r^{-\beta}f_{R_2}(r| r_1)  {\rm d}r\, r_1{\rm d}r_1 \nonumber\\
& +{\rm SAR_{\rm dl}}\,\frac{\bar{p}_{u,p}\, p_a\lambda_{\rm cu} r_{\rm c}^{2}}{4 } \int_{0}^{r_{\rm c}} \int_{0}^{r_{\rm c}+r_1} r^{-\beta}f_{R_2}(r| r_1)  {\rm d}r \frac{2r_{1}}{r_{\rm c}^2}{\rm d}r_1\nonumber\\
 &+{\rm SAR_{\rm dl}}\rho_{\rm b}  \frac{\lambda_{\rm b}G_{\rm b}}{2  (\beta-2)},\nonumber\\
&\bar{{\rm EI}}_{\rm p,2} = {\rm SAR_{\rm dl}}\,\frac{\bar{p}_{u,m}\, \lambda_{\rm c}  p_a\lambda_{\rm cu}\pi r_{\rm c}^{2}}{2  } \int_{0}^{\infty}\int_{0}^{r_{\rm c}+r_1} r^{-\beta}f_{R_2}(r| r_1)  {\rm d}r\, r_1{\rm d}r_1\nonumber\\
 & +{\rm SAR_{\rm dl}}\,\rho_{\rm b} \,\frac{\lambda_{\rm b}G_{\rm b}}{\pi   (\beta-2)} + \frac{{\rm SAR_{\rm dl}}}{4\pi}\int_{0}^{r_{\rm c}}\bigg(\rho_{\rm b}G_{\rm b}  r_{1}^{-\beta}+  \nonumber\\
&\frac{\bar{p}_{u,m}p_a\lambda_{\rm cu}\pi r_{\rm c}^{2}}{ }\int_{0}^{r_{\rm c}+r_1} r^{-\beta}f_{R_2}(r| r_1)  {\rm d}r\bigg) \frac{2r_{1}}{r_{\rm c}^2}{\rm d}r_1 .\label{eq_mean_EI_t_MCP}
\end{align}
\end{small}
\end{theorem}
\begin{IEEEproof}
Follows a similar way as PPP user model, thus omitted here.
\end{IEEEproof}

\subsubsection{Active user}
Compared to the Laplace transform of $W_{\rm u}$ of PPP user model, as well as the passive user case, the Laplace transform of $W_{\rm u}$ in the case of MCP  user  model is slightly different. This is because of the number of active intra-cluster uplink users: the total number of active uplink users in the same cluster is a Poisson random variable with parameter $n\sim {\rm Exp}(p_a\lambda_{\rm cu} \pi r_{\rm c}^2)$, the intra-cluster EMF exposure  $W_{u,c_0}$ is from all the cluster users except the active user. With that being said, $W_{u,c_0}= \sum_{k = 0}^{n-1}\frac{p_{u,x_{k}}H_{x_{k}}}{4\pi}D_{x_{k}}^{-\beta}$. 
The Laplace transform of $W_{\rm u,a,\{1,2\}}$ of the active user is given in the following lemma.

\begin{lemma}[Laplace Transform of $W_{\rm u,a,\{1,2\}}$, Scenario 1 \& 2] \label{lemma_Wu_reference_MCP}
In Scenario 1, the Laplace transform of $W_{u,a,1}$ of the active user is given by
\begin{align}
		&\mathcal{L}_{W_{\rm u,a,1}}(s)\approx  \exp\biggl\{-2\pi\lambda_{\rm c}\int_{0}^{\infty}1-\int_{0}^{\infty}\exp\bigg(-p_a\lambda_{\rm cu}\pi r_{\rm c}^{2}\nonumber\\
	&\int_{0}^{r_{\rm c}+r_1}\frac{f_{R_2}(r| r_1)}{1+(s p(z))^{-1} (4\pi) r^{\beta}}{\rm d}r\bigg)f_{R_{\rm u}}(z){\rm d}zr_1{\rm d}r_1\biggr\}\nonumber\\
	&\times\int_{0}^{r_{\rm c}}\int_{0}^{\infty} \frac{\exp(-p_a\lambda_{\rm cu}\pi r_{\rm c}^{2})}{\int_{0}^{r_{\rm c}+r_1}\frac{f_{R_2}(r| R_{\rm u})}{1+s p(z) (4\pi)^{-1} r^{-\beta}}{\rm d}r}\bigg(\exp\bigg(p_a\lambda_{\rm cu}\pi r_{\rm c}^{2}\nonumber\\
	&\int_{0}^{r_{\rm c}+r_1}\frac{f_{R_2}(r| r_1)}{1+s p(z) (4\pi)^{-1} r^{-\beta}}{\rm d}r\bigg)-1\bigg)f_{R_{\rm u}}(z){\rm d}z \frac{2 r_1}{r_{\rm c}^{2}}{\rm d}r_1.
\end{align}
In Scenario 2, the Laplace transform of $W_{\rm u}$ of the passive user is given by
\begin{small}
\begin{align}
&\mathcal{L}_{W_{\rm u,a,2}}(s)\approx  \exp\biggl\{-2\pi\lambda_{\rm c}\int_{0}^{\infty}1-\exp\bigg(-p_a\lambda_{\rm cu}\pi r_{\rm c}^{2}\int_{0}^{r_{\rm c}+r_1}\nonumber\\
	&\frac{f_{R_2}(r| r_1)}{1+(s \bar{p}_{u,m})^{-1} (4\pi) r^{\beta}}{\rm d}r\bigg)r_1{\rm d}r_1\biggr\} \int_{0}^{r_{\rm c}}\frac{\exp(-p_a\lambda_{\rm cu}\pi r_{\rm c}^{2})}{\int_{0}^{r_{\rm c}+r_1}\frac{f_{R_2}(r| r_1)}{1+s \bar{p}_{u,m}(4\pi)^{-1} r^{-\beta}}{\rm d}r}\nonumber\\
	&\bigg(\exp\bigg(p_a\lambda_{\rm cu}\pi r_{\rm c}^{2}\int_{0}^{r_{\rm c}+r_1}\frac{f_{R_2}(r| r_1)}{1+s \bar{p}_{u,m} (4\pi)^{-1} r^{-\beta}}{\rm d}r\bigg)-1\bigg) \frac{2 r_1}{r_{\rm c}^{2}}{\rm d}r_1.
\end{align}
\end{small}
\end{lemma}
\begin{IEEEproof}
See Appendix \ref{app_Wu_reference_MCP}.
\end{IEEEproof}

Obtained the Laplace transform of $W_{\rm u,a,\{1,2\}}$, we are able to derive the CDF and mean of ${\rm EI}_{\rm a}$, which  are given in the following two theorems.
\begin{theorem}[CDF and Mean of ${\rm EI}_{\rm a}$, Scenario 1 \& 2] \label{theore_CDF_EI_r_MCP}
The CDF of the total passive exposure, including exposure from BSs and active uplink users is given by
\begin{align}
&F_{{\rm EI_{a,\{1,2\}}}}(w) = \frac{1}{2}-\frac{1}{2j\pi}\int_{0}^{\infty}\int_{0}^{\infty}\frac{1}{ t}[\exp(-jtw)\mathcal{L}_{\rm EI_a}(-jt,r)\nonumber\\
&-\exp(jtw)\mathcal{L}_{\rm EI_a}(jt,r)]f_{R_{\rm u}^{\prime\prime}}(r){\rm d}r{\rm d}t,\nonumber
\end{align}
in which $f_{R_{\rm u}^{\prime\prime}}(r) = f_{R_{\rm u}}(r)$ in the case of Scenario 1 and $f_{R_{\rm u}^{\prime\prime}}(r) = f_{R_{\rm u}^{\prime}}(r)$ in the case of Scenario 2,
\begin{small}
\begin{align}
 &\mathcal{L}_{\rm EI_{a,\{1,2\}}}(s,R_{\rm u}^{\prime\prime}) =\mathcal{L}_{\{W_{\rm b,a},W_{\rm b,a}^{\prime}\}}(s\,{\rm SAR}_{\rm dl},R_{\rm u}^{\prime\prime})\nonumber\\
&\times\mathcal{L}_{W_{u,a,\{1,2\}}}(s\,{\rm SAR}_{\rm dl})\exp\bigg(-s\,{\rm SAR}_{\rm ul}\max(p_{\rm max},\rho_{\rm u} R_{\rm u}^{\prime\prime\alpha\eta})\bigg)\nonumber\\
&\times\bigg(\frac{1}{1+s{\rm SAR}_{\rm dl} \rho_{\rm b} G_{\rm b} R_{\rm u}^{\prime\prime-\beta}(4\pi)^{-1}}\bigg),\nonumber\\
&\mathcal{L}_{W_{\rm b,a}^{\prime}}(s,R_{\rm u}^{\prime\prime}) =  \mathcal{L}_{W_{\rm b,a}}(s\,{\rm SAR}_{\rm dl},R_{\rm u}^{\prime\prime})\exp\bigg(-s\,{\rm SAR}_{\rm dl}\frac{\rho_b G_{\rm b}}{4\pi}R_{u}^{\prime\prime-\beta}\bigg).
\end{align}
\end{small}
The mean of the total passive exposure, including exposure from BSs and active uplink users is given by
\begin{align}
&\bar{{\rm EI}}_{\rm a,\{1,2\}} = \int_{0}^{\infty}\bigg({\rm SAR_{\rm dl}}\,G_{\rm b}\rho_{\rm b}\,  (4\pi)^{-1}\bigg(r^{-\beta}+\frac{2\pi\lambda_{\rm b}G_{\rm b}}{\beta-2}r^{2-\beta}\bigg) \nonumber\\
&+{\rm SAR_{\rm ul}}\min(p_{\rm max},\rho_{\rm u} r^{\alpha\eta})\bigg)f_{R_{\rm u}^{\prime\prime}}(r) {\rm d}r +{\rm SAR_{\rm dl}}\{\bar{p}_{u,p},\bar{p}_{u,m}\} \nonumber\\
&\times\frac{\lambda_{\rm c}  p_a\lambda_{\rm cu}\pi r_{\rm c}^{2} }{2}\int_{0}^{\infty}\int_{0}^{r_{\rm c}+r_1} r^{-\beta}f_{R_2}(r| r_1)  {\rm d}r r_1{\rm d}r_1 \nonumber\\
&  +\frac{{\rm SAR_{\rm dl}}\{\bar{p}_{u,p},\bar{p}_{u,m}\}}{4\pi} \int_{0}^{r_{\rm c}} \bar{m} \int_{0}^{r_{\rm c}+r_1} r^{-\beta}f_{R_2}(r| r_1)  {\rm d}r \frac{2r_{1}}{r_{\rm c}^2}{\rm d}r_1,
\end{align}
in which $\bar{m} = \frac{1}{m}\exp(-m+1/m)-\exp(-m+1/m)+\exp(-m)$, in which $m = p_a\lambda_{\rm cu}\pi r_{\rm c}^{2}$, is the average number of active uses except the active user in the active user's cluster.
\end{theorem}
\begin{IEEEproof}
Follows the same lines as the proof of Theorem \ref{theorem_CDF_Mean_EI_r_PPP}, thus omitted here.
\end{IEEEproof}

\section{Coverage Probability}
In this section, we provide the analysis of the uplink coverage probability. 
At each resource block, there is only one active user in the PV cell formed by BSs (while all the active uplink users contribute to the EMF exposure, only one user in each PV cell contributes to the interference of uplink transmission) or in the user cluster that the BS serves. Recall that we study the uplink coverage probability at the tagged BS, the BS that serves the active user.
Let $R_i$ be the distance from an interfering uplink user to its serving BS, which should be strictly less than the distance from this uplink user to the tagged BS, $D$, since the user associates with the nearest BS. 
\begin{lemma}[Distribution of the Distance between an Intefering Uplink User and its Associated BS $R_i$]
The PDF of $R_i$ is given in \cite{haenggi2012stochastic}, as
\begin{align}
f_{R_i}(r|D) = \frac{2\pi\lambda_{\rm b} r\exp(-\lambda_{\rm b}\pi r^2)}{1-\exp(-\lambda_{\rm b}\pi D^2)}, \quad 0\leq r\leq D.
\end{align}
\end{lemma}

The density of uplink interferers observed from the tagged BS is well approximated by a non-homogeneous PPP with density $\lambda_{\rm b}(1-\exp(-\pi\lambda_{\rm b} d^2))$ \cite{singh2015joint},  in which $d$ denotes the distance between a user to the tagged BS. 

	\begin{lemma}[Laplace Transform of Interference and Noise]
The Laplace transform of the interference and noise is given by
\begin{align}
&\mathcal{L}_{I_1+\sigma^2}(s) =  \exp(-s\sigma^2)\exp\bigg(- \lambda_{\rm b}\int_{0}^{\infty} \int_{0}^{x}  \nonumber\\
&\frac{2\pi r\lambda_{\rm b}\exp(-\lambda_{\rm b}\pi r^2)}{1+s^{-1} \max(p_{\rm max}^{-1},\rho_{\rm u}^{-1} r^{-\alpha\eta}) x^{\alpha}G_{\rm b}^{-1}}{\rm d}r   x{\rm d}x\bigg),\nonumber\\
&\mathcal{L}_{I_2+\sigma^2}(s)=  \exp(-s\sigma^2)\exp\bigg(- \lambda_{\rm b}\int_{0}^{\infty}\int_{0}^{r_{\rm c}}\nonumber\\
&\frac{2r}{r_{\rm c}^2}\bigg[1- \frac{1}{1+s \min(p_{\rm max},\rho_{\rm u} r^{\alpha\eta}) x^{-\alpha}G_{\rm b}}\bigg] {\rm d}r x{\rm d}x\bigg).
\end{align}
in which $\mathcal{L}_{I_1+\sigma^2}(s)$ is the Laplace transform of PPP user model and MCP user model Scenario 1, and $\mathcal{L}_{I_2+\sigma^2}(s)$ is the Laplace transform of Scenario 2 in the case of MCP user model.
\end{lemma}
\begin{IEEEproof}
Similar to  \cite{singh2015joint}, thus omitted here.
\end{IEEEproof}
Finally, the uplink coverage probability is given in the following theorem.
\begin{theorem}[Uplink Coverage Probability]
Uplink coverage of PPP user model and MCP user model Scenario 1, is given by
\begin{align}
P_{\rm cov} &= 
\int_{0}^{\infty}f_{R_{\rm u}}(r)\mathcal{L}_{I_1+\sigma^2}(\gamma \max(p_{\rm max}^{-1},\rho_{\rm u}^{-1} r^{-\alpha\eta}) r^{\alpha}){\rm d}r,
\end{align}
and the uplink coverage probability of Scenario 2 in the case of MCP user model is given by,
\begin{align}
P_{\rm cov,2} &= 
\int_{0}^{r_{\rm c}}\frac{2r}{r_{\rm c}^2}\mathcal{L}_{I_2+\sigma^2}(\gamma \max(p_{\rm max}^{-1},\rho_{\rm u}^{-1} r^{-\alpha\eta}) r^{\alpha}){\rm d}r.
\end{align}

\end{theorem}
\begin{IEEEproof}
Uplink coverage probability is given by
\begin{align}
&P_{\rm cov} = 
\mathbb{P}\bigg({\rm SINR} > \gamma\bigg) = \mathbb{P}\bigg(G_{\rm b}\, p_{\rm u}\,H\, R_{\rm u}^{-\alpha} > \gamma(I+\sigma^2)\bigg)  \nonumber\\
&= \mathbb{E}[\exp(-G_{\rm b}^{-1}\,(I+\sigma^2)\gamma p_{\rm u}^{-1} R_{\rm u}^{\alpha})],
\end{align}
proof completes by taking the expectation over $R_{\rm u}$, while in the PPP user model and Scenario 1 of MCP user model $R_{\rm u}$ is the first contact distance of PPP, and in Scenario 2 of MCP user model $R_{\rm u}$ is the distance to the cluster center.
\end{IEEEproof}

\section{Numerical Results}
\label{sect_numerical}
Unless stated otherwise, we use the simulation parameters as listed in Table \ref{par_val}. In all the provided figures, we use markers for simulations and lines for analytical results.
\begin{table}\caption{Table of Parameters}\label{par_val}
\centering
\begin{center}
\resizebox{1\columnwidth}{!}{
	\renewcommand{\arraystretch}{1}
	\begin{tabular}{ {c} | {c} | {c}  }
		\hline
		\hline
		\textbf{Parameter} & \textbf{Symbol} & \textbf{Simulation Value}  \\ \hline
		Density of BSs & $\lambda_{\rm b}$ & $1$ BS/km$^{2}$ \\\hline
		Density of user clusters & $\lambda_{\rm c}$ & $1$ BS/km$^{2}$ \\\hline
		Antenna gain & $G_{\rm b}$ &  10 dBi\\\hline
		Frequency & $f_{ \{u,b\}}$ & $2600$ MHz\\\hline
		Power control parameters & $\rho_{\rm \{u,b\}}$ & 0.008 mW, 10 W\\\hline
		Maximum transmission power &  $p_{\rm max}$ &  200 mW\\\hline		
		User cluster radius &  $r_c$ &  100 m\\\hline
		Active probability &  $p_a$ &  1\\\hline
		Noise power & $\sigma^{\prime 2} $ & $10^{-12}$ W\\\hline
		SINR threshold & $\gamma$ & $20$ dB \\\hline
		path-loss parameters & $\alpha,\beta$ & $4,2.5$\\\hline
		SAR reduced in uplink/downlink & ${\rm SAR}_{\rm ul},{\rm SAR}_{\rm dl}$ & $0.0053$ $\frac{{\rm W}}{{\rm kg}}/{\rm W}$, 0.0042 $\frac{{\rm W}}{{\rm kg}}/\frac{{\rm W}}{{\rm m^2}}$ \cite{vermeeren2015low}
		\\\hline\hline
\end{tabular}}
\end{center}
\end{table}
Recall that ${\rm EI}_{\rm a}$ and ${\rm EI}_{\rm p}$, as well as their different components, are defined in Definition, \ref{def_EI} (\ref{eq_EI_p}) and (\ref{eq_EI_a}).

\begin{figure}
\centering
\includegraphics[width = 0.8\columnwidth]{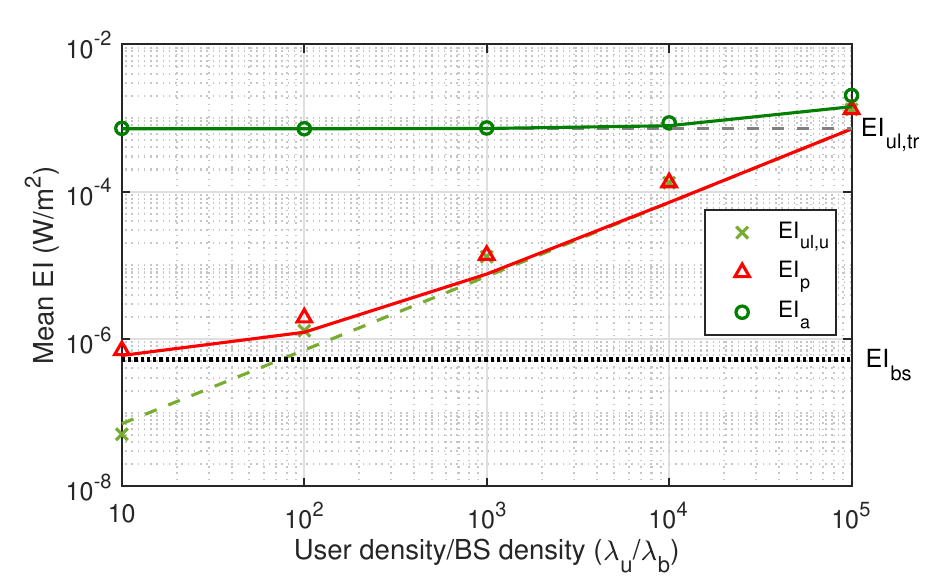}
\caption{ The impact of user density on the mean EI  and its different components: uplink users ${\rm EI_{ul,u}}$, BSs ${\rm EI_{bs}}$, and self-exposure from uplink transmission ${\rm EI_{ul,tr}}$ (only a component of active users' EI, ${\rm EI_a}$)), for the PPP user model with $\eta = 0.4$.}
\label{fig_ppp_diflambda_u1}
\end{figure}

\begin{figure}
\centering
\includegraphics[width = 0.8\columnwidth]{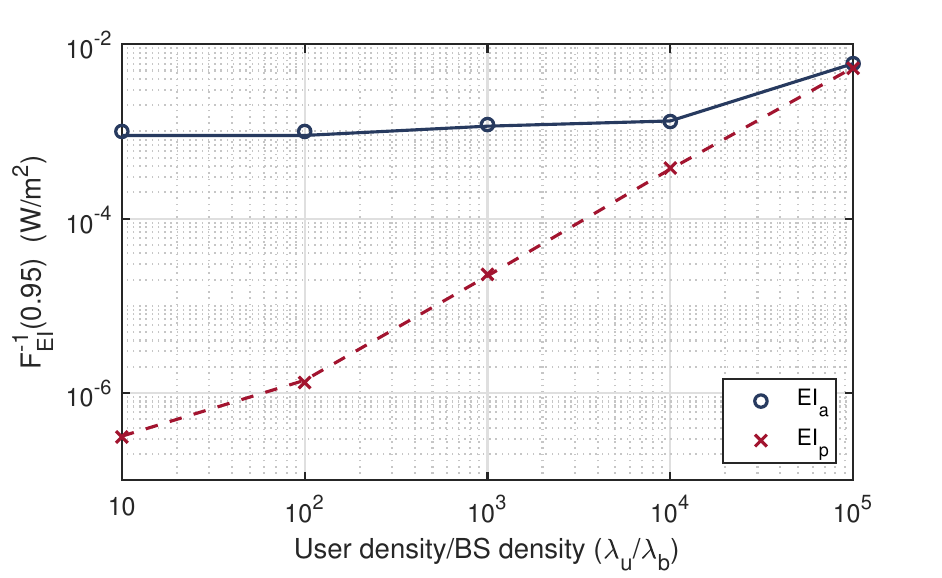}
\caption{The $95$-th-percentile of EI for passive and active users under different user density to BS density ratios with $\eta = 0.4$.}
\label{fig_ppp_diflambda_u2}
\end{figure}


In Fig.~\ref{fig_ppp_diflambda_u1}, we plot the mean of EI of the passive user (${\rm EI_{p}}$) and the active user (${\rm EI_{a}}$), respectively. Besides, we also plot the mean of EI from (i) BSs (${\rm EI_{bs}}$), (ii) self-exposure from uplink transmission (${\rm EI_{ul,tr}}$), and (iii) uplink transmission of other active uplink users (${\rm EI_{ul,u}}$). We show that when the ratio of active uplink user density to the BS density ($\lambda_{\rm u}/\lambda_{\rm b}$) is low, the dominant EMF exposure of the passive user and the active user comes from BSs and uplink transmission of the user itself, respectively.  With the increase of the ratio $\lambda_{\rm u}/\lambda_{\rm b}$, the EMF exposure from active uplink users increases and becomes of non-negligible magnitude, which is the intersection point in Fig.~\ref{fig_ppp_diflambda_u1}) compared to the EMF exposure from BSs starting from $\lambda_{\rm u}/\lambda_{\rm b} = 100$, and non-negligible compared to uplink transmission of an active user starting from $\lambda_{\rm u}/\lambda_{\rm b} = 10^5$, respectively.

In Fig.~\ref{fig_ppp_diflambda_u2}, we plot the $95$-th percentile of EI of the passive and active user, respectively. As the exposure in 5G networks is stochastic, this metric is essential to assess exposure performance in 5G networks where we need to limit it below some threshold. Similarly, we observe that the EMF exposure from other active uplink users becomes non-negligible at extremely high $\lambda_{\rm u}/\lambda_{\rm b}$, e.g., $\lambda_{\rm u}/\lambda_{\rm b} > 10^4$. 

\begin{figure}
\centering
\subfigure[]{\includegraphics[width = 0.8\columnwidth]{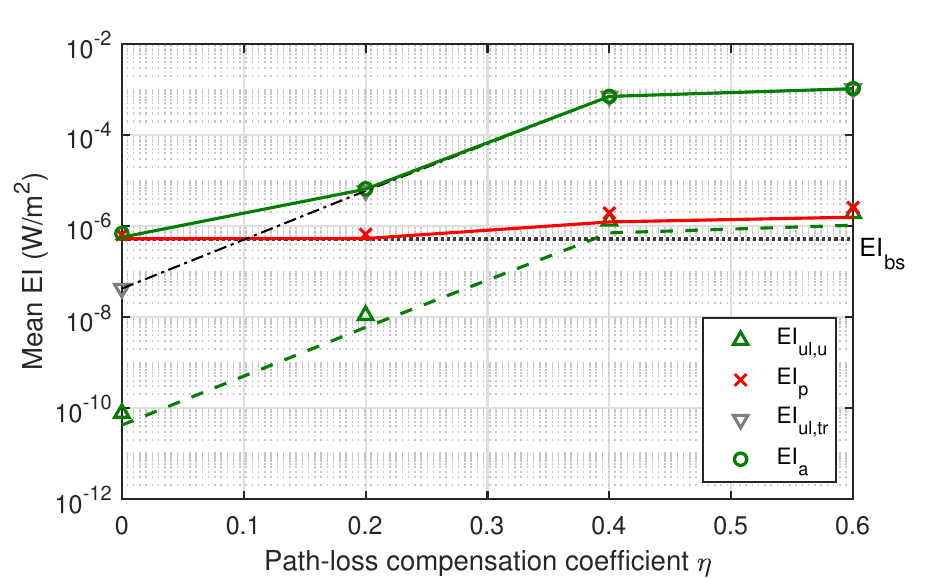}}
\subfigure[]{\includegraphics[width = 0.8\columnwidth]{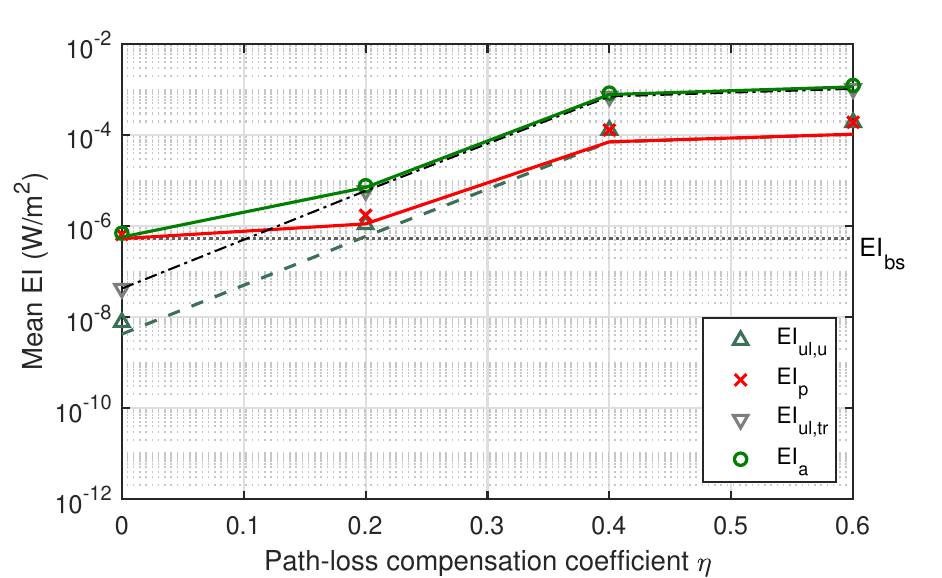}}		
\caption{Mean of EI of the active user ${\rm EI_{a}}$ and passive user ${\rm EI_{p}}$, and their different components: uplink users ${\rm EI_{ul,u}}$, BSs ${\rm EI_{bs}}$, and self-exposure from uplink transmission ${\rm EI_{ul,tr}}$, for different  $\eta$ at a fixed user density to BS density ratio for PPP user model \textbf{(a)}  $\lambda_{\rm u}/\lambda_{\rm b} = 100$, and \textbf{(b)}  $\lambda_{\rm u}/\lambda_{\rm b} = 10^4$.}
\label{fig_ppp_difeta}
\end{figure}
In Fig.~\ref{fig_ppp_difeta}, we plot the mean of EI under different uplink power control coefficients, $\eta$, for (a) a less crowded case with $\lambda_{\rm u}/\lambda_{\rm b} = 100$ and (b) an extreme case with $\lambda_{\rm u}/\lambda_{\rm b} = 10^4$. Note that in Fig.~\ref{fig_ppp_difeta} when $\eta = 0$, this implies the case where there is no path-loss compensation and all users transmit at the minimum transmit power, $\rho_{\rm u}$. In this case, the EMF exposure from BSs is much greater than that from the uplink transmission, either from the user itself or other active users. With the increase of $\eta$, the EMF exposure from uplink transmission increases dramatically at first and then increases slowly. This is due to the limitation of the transmit power. More precisely, when $\eta$ increases from $0.2$ to $0.4$, almost all the active users' transmit power increases to $p_{\rm max}$. On the other hand, when $\eta$ increases from $0.4$ to $0.6$, the transmit power of only a few active users (those located very close to the serving BS) increases to $p_{\rm max}$.

In the scenario of large user density to BS density ratio plotted in Fig.~\ref{fig_ppp_difeta}~(b), the mean of EI of the active user and passive user shows similar trends to the less crowded case plotted in Fig. \ref{fig_ppp_difeta} (a). Besides, we notice that comparing Fig.~\ref{fig_ppp_difeta}~(a) with Fig.~\ref{fig_ppp_difeta}~(b), while the mean of ${\rm EI_p}$ increases dramatically with the increase of $\lambda_{\rm u}/\lambda_{\rm b}$, the mean of ${\rm EI}_{\rm bs}$ and ${\rm EI}_{\rm ul,tr}$ keep the same.
Since ${\rm EI}_{\rm ul,tr}$ and ${\rm EI}_{\rm bs}$ are not a function of user density, the average distance to the nearest BS and the average transmit power do not change with $\lambda_{\rm u}$.



\begin{figure}
\centering
\includegraphics[width = 0.8\columnwidth]{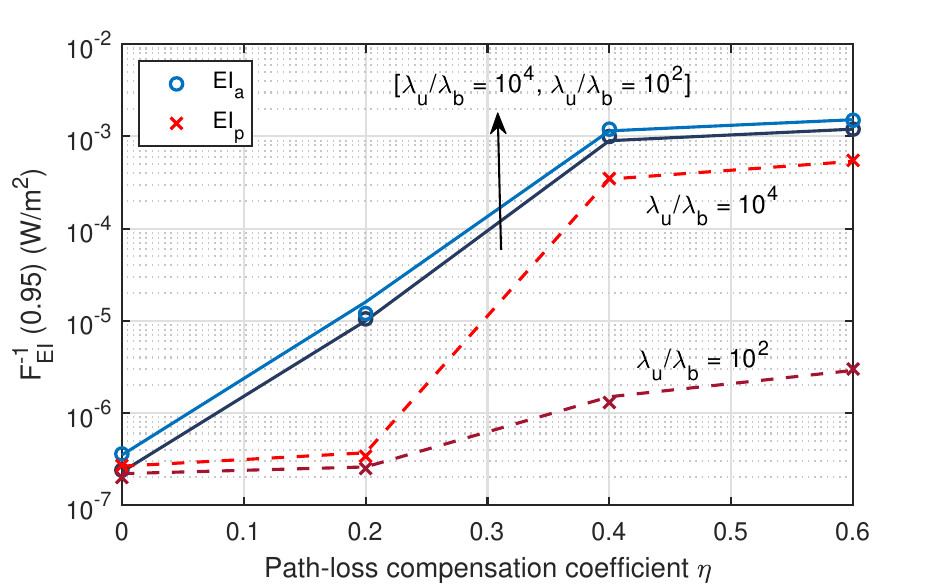}
\caption{ The $95$-th-percentile of EI for passive and active users  under  different  $\eta$ and at fixed user density to BS density ratio: $\lambda_{\rm u}/\lambda_{\rm b} = 100$ and $\lambda_{\rm u}/\lambda_{\rm b} = 10^4$.}
\label{fig_ppp_difeta_ext1}
\end{figure}

In Fig.~\ref{fig_ppp_difeta_ext1}, we show that at low values of $\eta$ the passive user and the active user experience a similar level of EMF exposure which is mainly from BSs, and at high values of $\eta$ the active user experiences a much higher level of EMF exposure because of its uplink transmission. With the increase of the user density to BS density ratio, the EMF exposure of the passive user increases dramatically while the exposure of the active user only changes slightly due to the fact that active exposure is much greater than passive exposure.

\begin{figure}
\centering
\includegraphics[width = 0.8\columnwidth]{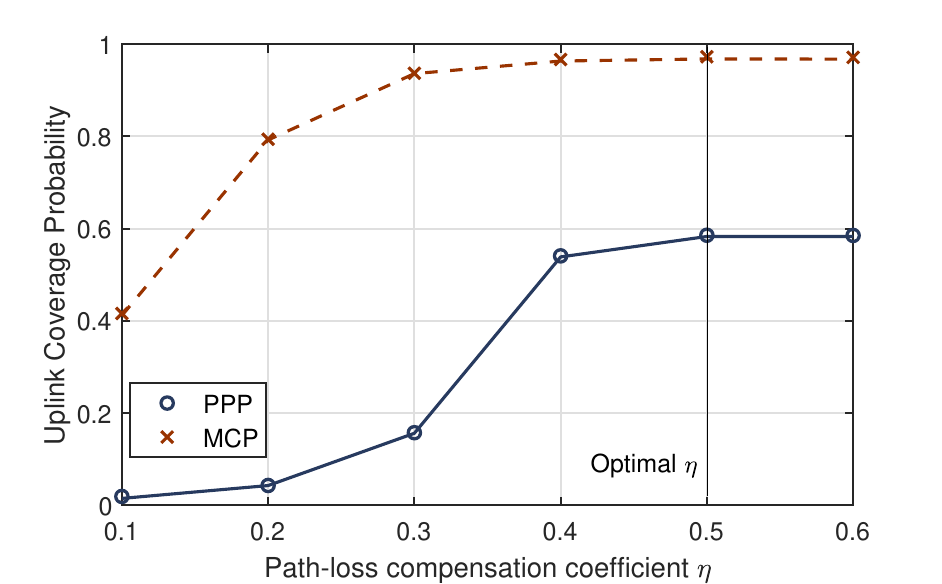}
\caption{Uplink coverage probability of PPP user model and MCP user model under different $\eta$.}
\label{fig_ppp_difeta_ext2}
\end{figure}
Fig.~\ref{fig_ppp_difeta_ext2} shows the uplink coverage probability under different $\eta$ for both PPP and MCP user models. We observe from the figure that there exists a value of $\eta$ that maximizes the coverage probability for both user models. This is because the increase in the interference caused by increasing $\eta$ becomes more significant than the increase in the quality of the uplink signal. Since we assume that BSs split the resources to serve the users to avoid intra-cell interference, uplink coverage probability is not a function of user densities, and it is only a function of $\eta$. With the increase of $\eta$, the transmit power increases, thus the coverage probability increases. However, EMF exposure to residents also increases. Therefore, there is a trade-off between good communication performance and EMF exposure. 

\begin{figure}
\centering
\includegraphics[width = 0.8\columnwidth]{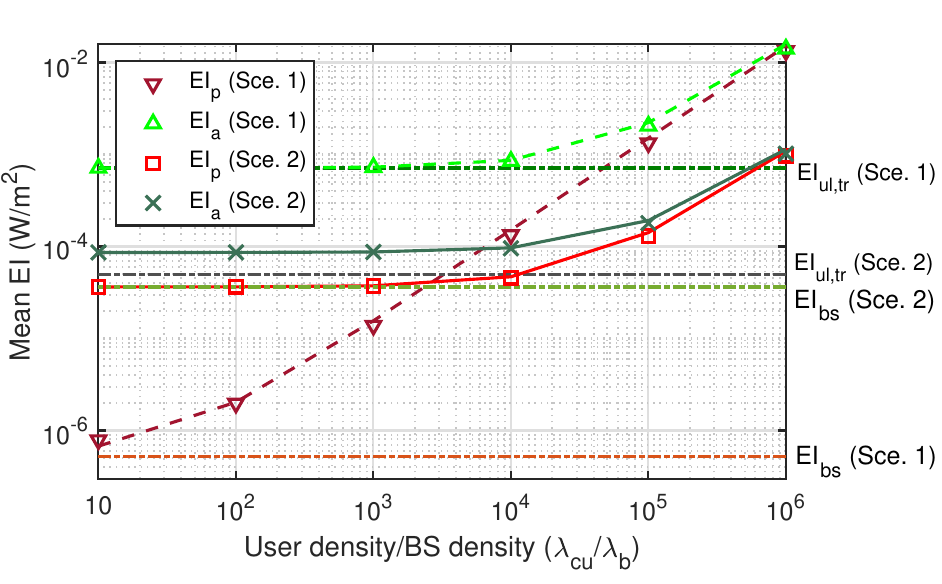}
\caption{The impact of user density to BS density ratio on the mean EI  and its different components: BSs ${\rm EI_{bs}}$ and self-exposure from uplink transmission ${\rm EI_{ul,tr}}$ (only a component of active users' EI), for the MCP user model with $\eta = 0.4$.}
\label{fig_mcp_diflambda_cu1}
\end{figure}

\begin{figure}[t!]
\centering
\includegraphics[width = 0.8\columnwidth]{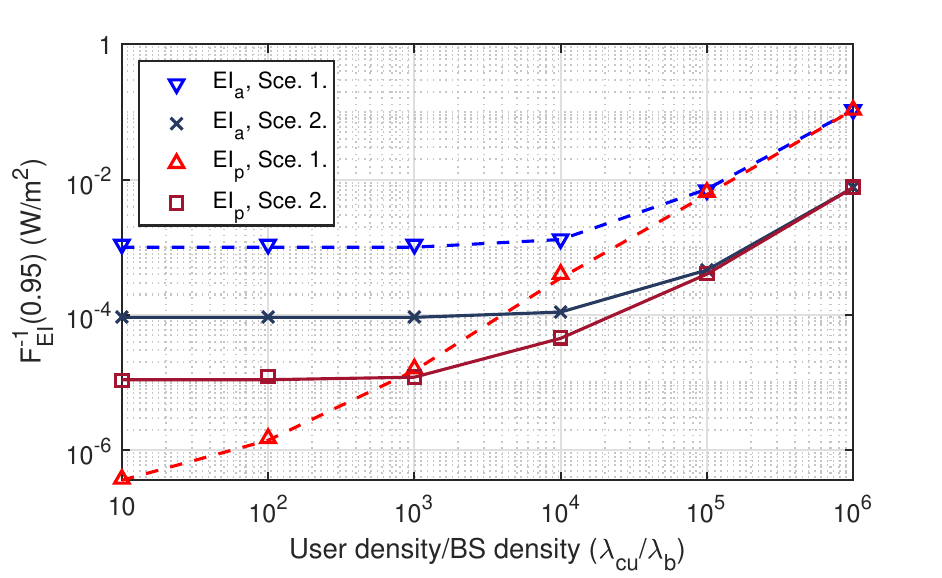}
\caption{ The $95$-th-percentile of EI for passive and active users for MCP user model in both scenarios, under different user density to BS density ratios and $\eta = 0.4$. }
\label{fig_mcp_diflambda_cu2}
\end{figure}

In Fig.~\ref{fig_mcp_diflambda_cu1}, we present the results for the mean of the EI in two different scenarios: Scenario 1, where user cluster centers are independently distributed according to a PPP, and Scenario 2, where user clusters are centered at the BSs.
Interestingly, we observe that when the user density to BS density ratio ($\lambda_{\rm cu}/\lambda_{\rm b}$) is low, both passive and active users in Scenario 2 experience higher EMF exposure compared to Scenario 1. 

In Fig.~\ref{fig_mcp_diflambda_cu2}, we present the results for the $95$-th percentile of the EI in the two aforementioned scenarios. We notice that in Scenario 1, the EMF exposure level increases gradually with the increase in $\lambda_{\rm cu}/\lambda_{\rm b}$. This finding suggests that when the density of users relative to the density of BSs is low, deploying BSs at the user cluster centers is not efficient. This is because, although placing a BS at the cluster center significantly reduces the transmit power of UE, it also increases exposure due to the high transmit power of the BS and the shorter distance between the BS and the users. However, if $\lambda_{\rm cu}/\lambda_{\rm b}$ reaches $10^3$, deploying BSs at the user cluster centers becomes beneficial.



\begin{figure}
\centering
\includegraphics[width = 0.8\columnwidth]{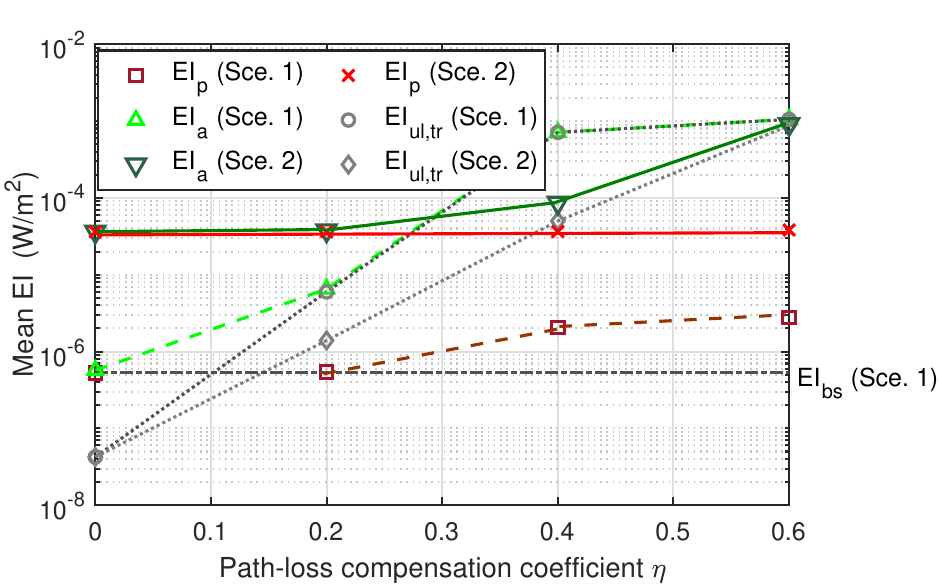}
\caption{The impact of user density on the mean EI  and its different components: BSs ${\rm EI_{bs}}$, and self-exposure from uplink transmission ${\rm EI_{ul,tr}}$ (only a component of active users' EI), for the MCP user model under  different $\eta$  and $\lambda_{\rm cu}/\lambda_{\rm b} = 100$.}
\label{fig_mcp_difeta1}
\end{figure}

\begin{figure}
\centering
\includegraphics[width = 0.8\columnwidth]{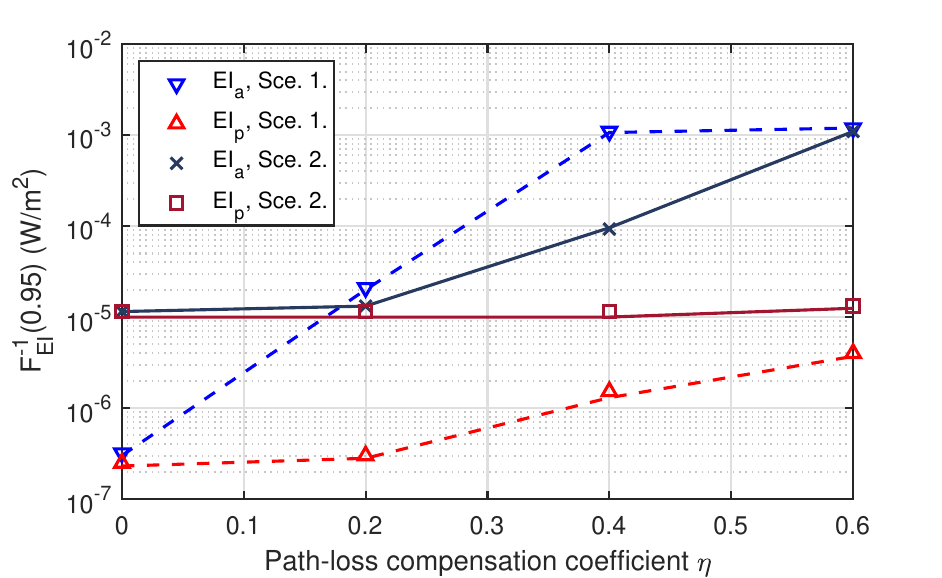}
\caption{The $95$-th-percentile of EI for passive and active users for MCP user model in both scenarios under different $\eta$  and $\lambda_{\rm cu}/\lambda_{\rm b} = 100$.}
\label{fig_mcp_difeta2}
\end{figure}

Fig.~\ref{fig_mcp_difeta1} shows the mean of EI under different $\eta$ for MCP user model. ${\rm EI_{ul,tr}} $ (Sce. 1) and ${\rm EI_{ul,tr}}$ (Sce. 2) start at the same value when $\eta = 0$ since the active user has the same transmit power $p_o = \rho_{\rm u}$ without path-loss compensation. As $\eta$ increases, ${\rm EI_{ul,tr}}$ (Sce. 1) becomes greater than ${\rm EI_{ul,tr}}$ (Sce. 2). This is because the active user requires a higher transmit power to compensate for the path-loss when the BS is located further away in Scenario 1 compared to Scenario 2. However, as $\eta$ continues to increase, ${\rm EI_{ul,tr}}$ reaches its maximum achievable value, limited by the maximum transmit power of the devices. 

In Fig.~\ref{fig_mcp_difeta2} we show the $95$-th-percentile of EI under different $\eta$ for MCP user model. We notice that while Scenario 1 is always beneficial for the passive user, it is only beneficial for the active users at low values of $\eta$.

\begin{figure}
\centering
\subfigure[]{\includegraphics[width =  0.8\columnwidth]{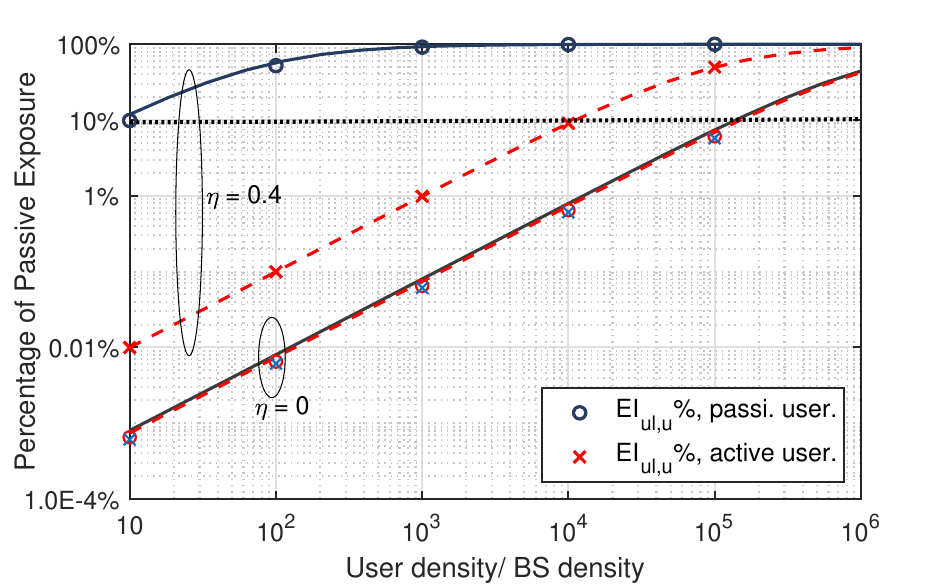}}
\subfigure[]{\includegraphics[width = 0.8\columnwidth]{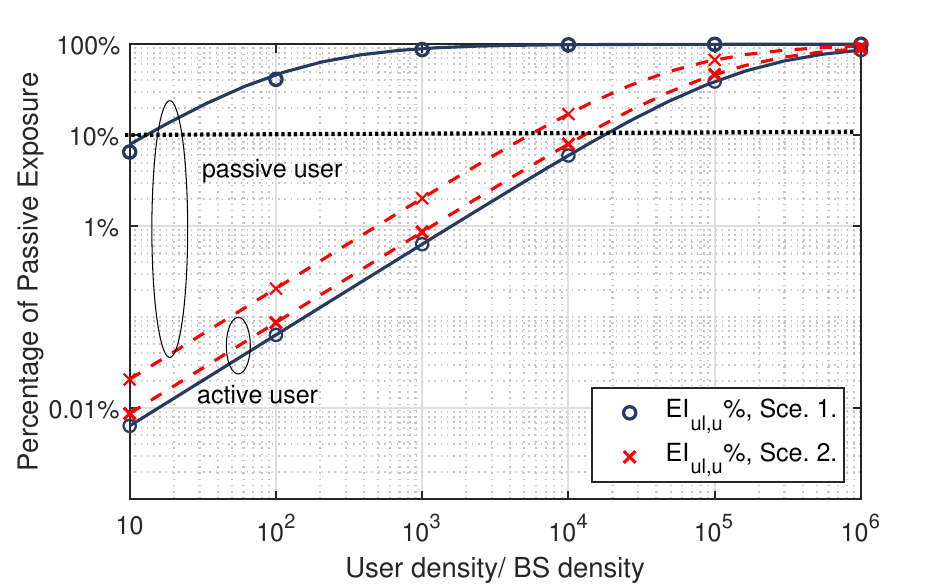}}	
\caption{  \textbf{(a)} Percentage of ${\rm EI_{ul,u}}$ to the total exposure in PPP user model and \textbf{(b)} Percentage of ${\rm EI_{ul,u}}$ to the total exposure in MCP user model.}
\label{fig_percentage}
\end{figure}
To further show the impact of the passive EMF exposure, in Fig.~\ref{fig_percentage},  we plot the percentage of ${\rm EI_{ul,u}}$ with respect to the total EMF exposure for both passive and active users, respectively. With the increase of the user density, the passive exposure from uplink users cannot be ignored if it contributes more than $10\%$ of the total exposure (dotted line in Fig.~\ref{fig_percentage}): in PPP user case when $\eta=0.4$, ${\rm EI_{ul,u}}$ cannot be ignored when $\lambda_{\rm u}/\lambda_{\rm b} \geq 10$ for the passive user while $\lambda_{\rm u}/\lambda_{\rm b} \geq  10^4$ for the active user. Interestingly, we notice that in MCP case, ${\rm EI_{ul,u}}$ has a higher impact on the active user when BSs are deployed at the cluster center compared to independent distribution. Even though deploying one BS at the cluster center reduces the ${\rm EI_{ul,u}}$, the overall exposure reduces in both active exposure and ${\rm EI_{ul,u}}$. Thus, ${\rm EI_{ul,u}}$ of Sce. 2 contributes more to overall exposure than in Sce. 1.

\section{Conclusion}
In this study, we introduced a novel stochastic geometry framework to analyze the EMF exposure experienced by both passive and active users in cellular network. Unlike conventional approaches that neglect the EMF exposure from other uplink users, our analysis accounted for this crucial aspect, leading to more accurate and realistic results. 
We showed that passive exposure (exposure from uplink transmission of other users) accounts for more than $10\%$ of the total exposure of the passive user even when the ratio between the user density and the BS density is as low as $10$ users per BS. On the other hand, for active users, passive exposure starts to exceed $10\%$ of the total exposure when the user density to the BS density ratio exceeds $10^4$ users per BS.
This research contributes to the understanding of EMF exposure in wireless networks and highlights the importance of considering interactions between users in the EMF exposure assessment.


\appendix
\subsection{Proof of Lemma \ref{lemma_Wu_passive_MCP}}\label{app_Wu_passive_MCP}
 In the case of the MCP user model, assuming the typical user is located at the origin. Therefore, we are conditioning on a typical cluster containing the origin, and the Laplace transform of $W_{\rm u}$ of a typical user is given by
\begin{small}
\begin{align}
&\mathcal{L}_{W_{\rm u}}(s) 
= \mathbb{E}_{W_{\rm u}}\bigg[\exp\bigg(-s\sum_{x_{u,ij}\in\Phi_{\rm cu}^{0'}}\frac{p_{u,x_{u,ij}} H_{x_{u,ij}} ||x_{u,ij}||^{-\beta}}{4\pi  }\bigg)\bigg]\nonumber\\
%
&\stackrel{(a)}{\approx}\mathbb{E}_{\Phi_{\rm c},p_{\rm u}}\bigg[\prod_{c_{i}\in\Phi_{\rm c}} \sum_{n = 0}^{\infty}\frac{\exp(-\lambda_{\rm cu}\pi r_{\rm c}^{2})(\lambda_{\rm cu}\pi r_{\rm c}^{2})^n}{n!}\bigg(\int_{0}^{r_{\rm c}+||c_i||} \nonumber\\
&\frac{f_{R_2}(r| ||c_i||)}{1+s p_{\rm u} (4\pi  )^{-1} r^{-\beta}}{\rm d}r\bigg)^{n}\bigg]\mathbb{E}_{R_{\rm u},p_{\rm u}}\bigg[\sum_{n = 0}^{\infty}\frac{\exp(-\lambda_{\rm cu}\pi r_{\rm c}^{2})(\lambda_{\rm cu}\pi r_{\rm c}^{2})^n}{n!} \nonumber\\
&\quad \times \bigg(\int_{0}^{r_{\rm c}+R_{\rm u}}\frac{f_{R_2}(r| R_{\rm u})}{1+s p_{\rm u} (4\pi  )^{-1} r^{-\beta}}{\rm d}r\bigg)^{n}\bigg]\nonumber\\
&\stackrel{(b)}{=}\mathbb{E}_{\Phi_{\rm c},p_{\rm u}}\bigg[\prod_{c_{i}\in\Phi_{\rm c}}\exp\bigg(-\lambda_{\rm cu}\pi r_{\rm c}^{2}\bigg(1-\int_{0}^{r_{\rm c}+||c_i||}\nonumber\\
&\quad\frac{f_{R_2}(r| ||c_i||)}{1+s p_{\rm u} (4\pi  )^{-1} r^{-\beta}}{\rm d}r\bigg)\bigg) \bigg]\mathbb{E}_{R_{\rm u},p_{\rm u}}\bigg[\exp\bigg(-\lambda_{\rm cu}\pi r_{\rm c}^{2}\bigg(1-\nonumber\\
&\quad\int_{0}^{r_{\rm c}+R_{\rm u}}\frac{f_{R_2}(r| R_{\rm u})}{1+s p_{\rm u} (4\pi  )^{-1} r^{-\beta}}{\rm d}r\bigg)\bigg)\bigg],
\end{align}	
\end{small}
in which the approximate sign comes from the assumption that users in one cluster have the same transmit power, step (a) follows computing the inter-cluster EMF and intra-cluster EMF separately, using the MGF of exponential random variables, computing the parent process and offspring process separately, and the i.i.d. property of each cluster and the number of active uplink users is a Poisson random variable $m\sim{\rm Exp}(\lambda_{\rm cu}\pi r_{\rm c}^2)$, and (b) follows from the Taylor expansion of an exponential function. 


Recall that we assume users in one cluster have the same transmit power in  scenario 1, therefore, $p_{\rm u}$ is a function of the distance from the cluster center to the nearest BS, and given by
\begin{small}
	\begin{align}
&\mathcal{L}_{W_{u,1}}(s)=\mathbb{E}_{\Phi_{\rm c}}\bigg[\prod_{c_{i}\in\Phi_{\rm c}}\int_{0}^{\infty}\exp\bigg(-\lambda_{\rm cu}\pi r_{\rm c}^{2}\bigg(1-\int_{0}^{r_{\rm c}+||c_i||}\nonumber\\
&\frac{f_{R_2}(r| ||c_i||)}{1+s p(z) (4\pi  )^{-1} r^{-\beta}}{\rm d}r\bigg)\bigg)f_{R_{\rm u}}(z){\rm d}z \bigg] \mathbb{E}_{R_{\rm u}}\bigg[\int_{0}^{\infty}\exp\bigg(-\lambda_{\rm cu}\nonumber\\
& \times\pi r_{\rm c}^{2}\bigg(1-\int_{0}^{r_{\rm c}+R_{\rm u}}\frac{f_{R_2}(r| R_{\rm u})}{1+s p(z) (4\pi  )^{-1} r^{-\beta}}{\rm d}r\bigg)\bigg)f_{R_{\rm u}}(z){\rm d}z\bigg],
\end{align}
\end{small}
recall that $p(z) = \min(p_{\rm max},\rho_{\rm u}z^{\eta\alpha})$ and proof completes by using the PGFL of PPP and taking the expectation over $R_{\rm u}$. In scenario 2, user clusters are centered at the BS, hence, the transmit power of users is a constant.
\begin{align}
&\mathcal{L}_{W_{u,2}}(s) =\mathbb{E}_{\Phi_{\rm c}}\bigg[\prod_{c_{i}\in\Phi_{\rm c}} \exp\bigg(-\lambda_{\rm cu}\pi r_{\rm c}^{2}\bigg(1-\int_{0}^{r_{\rm c}+||c_i||}\nonumber\\
&\quad \frac{f_{R_2}(r| ||c_i||)}{1+s \bar{p}_{u,m} (4\pi  )^{-1} r^{-\beta}}{\rm d}r\bigg)\bigg) \bigg]\mathbb{E}_{R_{\rm u}}\bigg[ \exp\bigg( -\lambda_{\rm cu}\pi r_{\rm c}^{2}\nonumber\\
&\quad\times\bigg(1-\int_{0}^{r_{\rm c}+R_{\rm u}}\frac{f_{R_2}(r| R_{\rm u})}{1+s \bar{p}_{u,m} (4\pi  )^{-1} r^{-\beta}}{\rm d}r\bigg)\bigg) \bigg],
\end{align}
recall that $\bar{p}_{u,m} = \int_{0}^{\infty}\min(p_{\rm max},\rho_{\rm u}z^{\eta\alpha})\frac{2 z}{r_{\rm c}^2}{\rm d}z$, and similarly to $\mathcal{L}_{W_{u,1}}(s)$, proof completes by using the PGFL of PPP and taking the expectation over $R_{\rm u}$.

\subsection{Proof of Lemma \ref{lemma_Wu_reference_MCP}}\label{app_Wu_reference_MCP}
Recall that as for the active user, while the expression of inter-cluster EMF exposure keeps the same as passive user case, the expression of the intra-cluster EMF exposure is slightly different since the number of active uplink user within the reference's user cluster is $n-1$ (in which $n$ is the total number of the active uplink user). Therefore, the Laplace transform of $W_{\rm u,a}$ of the active user is given by
\begin{small}
\begin{align}
&\mathcal{L}_{W_{\rm u,a}}(s) =\mathbb{E}_{\Phi_{\rm c},p_{\rm u}}\bigg[\prod_{c_{i}\in\Phi_{\rm c}} \sum_{n = 0}^{\infty}\frac{\exp(-\lambda_{\rm cu}\pi r_{\rm c}^{2})(\lambda_{\rm cu}\pi r_{\rm c}^{2})^n}{n!} \bigg(\int_{0}^{r_{\rm c}+||c_i||}\nonumber\\
&\frac{f_{R_2}(r| ||c_i||)}{1+s p_{u,c_i} (4\pi)^{-1} r^{-\beta}}{\rm d}r\bigg)^{n}\bigg]\mathbb{E}_{R_{\rm u},p_{\rm u}}\bigg[\sum_{n = 1}^{\infty}\frac{\exp(-\lambda_{\rm cu}\pi r_{\rm c}^{2})(\lambda_{\rm cu}\pi r_{\rm c}^{2})^n}{n!} \nonumber\\
&\times\bigg(\int_{0}^{r_{\rm c}+R_{\rm u}}\frac{f_{R_2}(r| R_{\rm u})}{1+s p_{u,c_0} (4\pi)^{-1} r^{-\beta}}{\rm d}r\bigg)^{n-1}\bigg],
%
\end{align}
\end{small}
in the case of Scenario 1,
\begin{small}
\begin{align}
&\mathcal{L}_{W_{\rm u,a,1}}(s)	=
\mathbb{E}_{\Phi_{\rm c}}\bigg[\prod_{c_{i}\in\Phi_{\rm c}}\int_{0}^{\infty}\exp\bigg(-\lambda_{\rm cu}\pi r_{\rm c}^{2}\bigg(1-\int_{0}^{r_{\rm c}+||c_i||}f_{R_{\rm u}}(z)\nonumber\\
&  \frac{f_{R_2}(r| ||c_i||)}{1+s p(z) (4\pi)^{-1} r^{-\beta}}{\rm d}r\bigg)\bigg) {\rm d}z\bigg]\mathbb{E}_{R_{\rm u}}\bigg[\int_{0}^{\infty}\frac{\exp(-\lambda_{\rm cu}\pi r_{\rm c}^{2})}{\int_{0}^{r_{\rm c}+R_{\rm u}}\frac{f_{R_2}(r| R_{\rm u})}{1+s p(z) (4\pi)^{-1} r^{-\beta}}{\rm d}r}\nonumber\\
&\bigg(\exp\bigg(\lambda_{\rm cu}\pi r_{\rm c}^{2}\int_{0}^{r_{\rm c}+R_{\rm u}}\frac{f_{R_2}(r| R_{\rm u})}{1+s p(z) (4\pi)^{-1} r^{-\beta}}{\rm d}r\bigg)-1\bigg)f_{R_{\rm u}}(z){\rm d}z\bigg],
\end{align}
\end{small}
in the case of Scenario 2,
\begin{align}
&\mathcal{L}_{W_{\rm u,a,2}}(s)	=
\mathbb{E}_{\Phi_{\rm c}}\bigg[\prod_{c_{i}\in\Phi_{\rm c}}\exp\bigg(-\lambda_{\rm cu}\pi r_{\rm c}^{2}\bigg(1-\int_{0}^{r_{\rm c}+||c_i||}\nonumber\\
&\frac{f_{R_2}(r| ||c_i||)}{1+s \bar{p}_{u,m} (4\pi)^{-1} r^{-\beta}}{\rm d}r\bigg)\bigg) \bigg]\mathbb{E}_{R_{\rm u}}\bigg[\frac{\exp(-\lambda_{\rm cu}\pi r_{\rm c}^{2})}{\int_{0}^{r_{\rm c}+R_{\rm u}}\frac{f_{R_2}(r| R_{\rm u})}{1+s \bar{p}_{u,m}(4\pi)^{-1} r^{-\beta}}{\rm d}r}\nonumber\\
&\times\bigg(\exp\bigg(\lambda_{\rm cu}\pi r_{\rm c}^{2}\int_{0}^{r_{\rm c}+R_{\rm u}}\frac{f_{R_2}(r| R_{\rm u})}{1+s \bar{p}_{u,m} (4\pi)^{-1} r^{-\beta}}{\rm d}r\bigg)-1\bigg)\bigg],
\end{align}
proof completes by using the PGFL of PPP and taking the expectation of $R_{\rm u}$.

\bibliographystyle{IEEEtran}
\bibliography{Rep15}
\end{document}